\documentclass[useAMS,usenatbib]{mn2e}

\usepackage[T1]{fontenc}
 \usepackage{pslatex}
\usepackage{amsmath}
\usepackage{amsfonts}
\usepackage{color}
\usepackage{url}
\usepackage{graphicx}
\usepackage{subfigure}
\usepackage{amssymb}
\usepackage{soul}
\usepackage{booktabs}
\usepackage{array}
\usepackage{textcomp}

{\newif\ifnotend
\notendtrue
\def\veclist{ABCDEFGHIJKLMNOPQRSTUVWXYZabcdefghijklmnopqrstuvwxyz.}
\def\top#1#2.{#1}
\def\tail#1#2.{#2.}
\loop\expandafter\xdef\csname v\expandafter\top\veclist\endcsname%
{{\noexpand\bf\expandafter\top\veclist}}
\edef\veclist{\expandafter\tail\veclist}
\if\veclist.\notendfalse\fi\ifnotend\repeat}
\def\e{{\rm e}}

\mathchardef\mhyphen="2D

\title[Long GRBs and Type Ib/c SNe
]{
A Common Central Engine for Long Gamma Ray Bursts and Type Ib/c Supernovae?
}
\author[Sobacchi, Granot, Bromberg \& Sormani]{E. Sobacchi$^{1,2}$\thanks{Contact email: sobacchi@post.bgu.ac.il}, J. Granot$^2$, O. Bromberg$^3$ \& M. C. Sormani$^4$\\
$^1$ Physics Department, Ben-Gurion University, P.O. Box 653, Beer-Sheva 84105, Israel\\
$^2$ Department of Natural Sciences, The Open University of Israel, 1 University Road, P.O. Box 808, Raanana 4353701, Israel\\
$^3$  The Raymond and Beverly Sackler School of Physics and Astronomy, Tel Aviv University, Tel Aviv 69978, Israel\\
$^4$ Institute for Theoretical Astrophysics, Zentrum f\"{u}r Astronomie der Universit\"{a}t Heidelberg, Albert-\"{U}berle-Str. 2, 69120 Heidelberg, Germany
}

\begin{document}

\date{}

\newcommand{\degree}{\ensuremath{^\circ}}

\maketitle

\begin{abstract}
Long-duration, spectrally-soft Gamma-Ray Bursts (GRBs) are associated with Type Ic Core Collapse (CC) Supernovae (SNe), and thus arise from the death of massive stars. In the collapsar model, the jet launched by the central engine must bore its way out of the progenitor star before it can produce a GRB. Most of these jets do not break out, and are instead ``choked'' inside the star, as the central-engine activity time, $t_{\rm e}$, is not long enough.
Modelling the long-soft GRB duration distribution assuming a power-law distribution for their central-engine activity times, $\propto t_{\rm e}^{-\alpha}$ for $t_{\rm e}>t_{\rm b}$,
we find a steep distribution ($\alpha\sim4$) and {\it a typical GRB jet breakout time of }$t_{\rm b}\sim 60\text{ s}$ {\it in the star's frame. The latter suggests the presence of a low-density, extended envelope surrounding the progenitor star}, similar to that previously inferred for low-luminosity GRBs.
Extrapolating the range of validity of this power law below what is directly observable, to  $t_{\rm e}<t_{\rm b}$, by only a factor of $\sim4 \mhyphen 5$ produces enough events to account for all Type Ib/c SNe. Such extrapolation is necessary to avoid fine-tuning the distribution of central engine activity times with the breakout time, which are presumably unrelated.
We speculate that {\it central engines launching relativistic jets may operate in all Type Ib/c SNe}. In this case, the existence of a common central engine would imply that (i) the jet may significantly contribute to the energy of the SN; (ii) various observational signatures, like the asphericity of the explosion, could be directly related to jet's interaction with the star.
\end{abstract}

%%%%%%%%%%%%%%%%%%%%%%%%%%%%%%%%%%%%%%%%%

\section{Introduction}
\label{sec:introduction}

GRBs are the most luminous explosions in the Universe. They are divided into two classes according to the duration and spectral hardness of their prompt gamma-ray emission \citep{Kouveliotou1993}: 
(i) Long-duration ($\gtrsim2\;$s) soft-spectrum
GRBs (e.g. \citealt{WoosleyBloom2006,KumarZhang2015}). These are found in star-forming regions and are associated with broad-lined
Type Ic supernovae, implying a massive star progenitor, which is most likely low-metallicity and rapidly rotating near this cataclysmic end of its life, and lives in a gas-rich environment not far from its birthplace; (ii) Short-duration ($\lesssim2\;$s) hard-spectrum GRBs (e.g. \citealt{Nakar2007}). These are thought to arise from the merger of a binary neutron star system (or a neutron star and a stellar-mass black hole) that emits gravitational waves as it inspirals and coalesces, producing a central-engine driven jet. Such systems live in low-density environments, possibly after experiencing a prior supernova kick that pushed them into the outskirts of their host galaxies. A third subclass, whose importance was realized only relatively recently
\citep{Soderberg2006,Campana2006,Bromberg2011b,NakarSari2012}, involves low-luminosity GRBs ({\it ll}GRBs), whose overall isotropic equivalent radiated energy is $E_{\rm\gamma,iso}\lesssim 10^{49}\;$erg. These also typically have a smooth, single-peaked light curve, and a $\nu F_\nu$ spectrum that typically peaks at a lower than average photon energy (usually $E_{\rm p} \lesssim 100\;$keV). Although they are rarely observed because of their low luminosity, they are more numerous than regular long GRBs in terms of rate per unit volume, and they most likely do not arise from the same emission mechanism (e.g. \citealt{Bromberg2011b,NakarSari2012}).

In order to produce a long-soft GRB (class (i) above), the central engine must drive a strong relativistic jet that bores its way through the stellar envelope and produces the GRB well outside of the progenitor star. The model providing the theoretical framework to interpret the SN-GRB association is known as the collapsar model \citep{MacfadyenWoosley1999, Macfadyen2001}. According to this model, following the core collapse of a massive star, a bipolar jet is launched from the center of the star. The central engine could involve either a jet launched through rapid accretion onto a newly formed black hole, or an MHD outflow from a newly formed rapidly-rotating, highly-magnetized neutron star (millisecond magnetar). In both cases the outflow is collimated into a narrow bipolar jet due to its interaction with the stellar envelope. The jet drills through the stellar envelope and breaks out of the surface before producing the observed gamma rays. 

\citet{Bromberg2012} have shown that such a ``jet in a star'' scenario naturally predicts the existence of a plateau in the GRB duration distribution. After correcting for cosmological time dilation, the upper end of this plateau (which they found to be at $\sim10\;$s) is in good agreement with the expected breakout time from a compact progenitor star ($\sim10\mhyphen 15\text{ s}$ for a hydrodynamic jet; \citealt{Bromberg2011,Bromberg2015}). However, as already noted by \citet{Bromberg2012}, the end of the plateau provides only a lower limit to the true breakout time. This motivates a new analysis, which includes a fit to the full expected functional form of the GRB duration distribution, and can determine the breakout time rather than only set a lower bound on it.

On the other hand, the mechanism driving the explosion of CC SNe is still unclear. The most popular neutrino-driven scenario \citep{BetheWilson1985} faces several difficulties to reproduce the observed events (see for example \citealt{Papish2015}). Interestingly, several authors (e.g. \citealt{Khokhlov1999,MaedaNomoto2003,Couch2009}) found that these explosions could potentially be powered by Newtonian, supersonic jets as well.

Since both types of events -- long-soft GRBs and CC SNe -- are associated with the death of massive stars, it is natural to ask whether or how often ordinary CC SNe also posses a central engine launching a jet, similar to that operating in long-soft GRBs. To answer this question, we use the simplest phenomenological model for the central engine activity times that reproduces the observed duration distribution of long-soft GRBs; fitting the GRB duration distribution through this model results in a breakout time of $\sim 60\text{ s}$ in the rest frame of the star, which is unusually long for a compact progenitor. We then compare (i) direct estimates of observed rates of long-soft GRBs, $R_{\rm GRB}$, with (ii) estimates of $R_{\rm GRB}$ derived from the observed rate of CC SNe of Type Ib/c. Coupling these results with our phenomenological model, we obtain constraints on the fraction of SNe Ib/c that harbour central engines launching a jet, and show that this fraction may be consistent with unity.

The paper is organised as follows. Section \ref{sec:collapsar} briefly reviews the most relevant aspects of the collapsar model and presents our simple phenomenological model for the duration distribution of central engine activity times and the associated fit to the overall GRB duration distribution. In Section \ref{sec:constraints} constraints are derived on the fraction of ordinary SNe Ib/c that harbour a central engine similar to that of long-soft GRBs. Section \ref{sec:discussion} discusses the implications of these results for (i) the dynamics and the geometry of CC SNe explosions, under the hypothesis that a significant fraction of them has a central engine launching a jet; (ii) the structure of the progenitor star of long-soft GRBs, as implied by the long breakout time that we are finding. Finally, Section \ref{sec:concl} summarises our conclusions.

\section{The collapsar model for long GRB\MakeLowercase{s}}
\label{sec:collapsar}

\begin{figure*}{\vspace{3mm}} 
\centering
\includegraphics[width=0.99\textwidth]{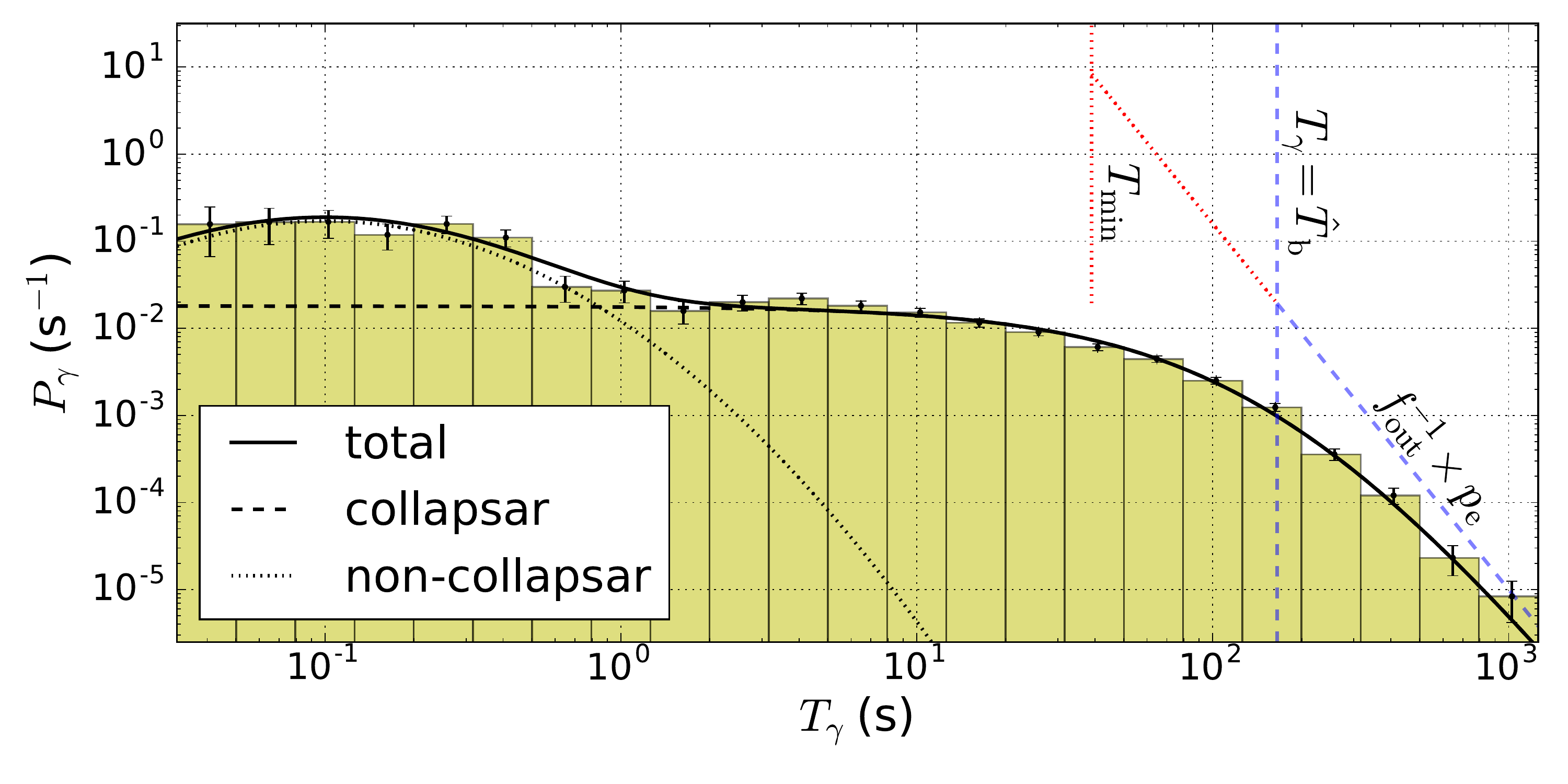}
\caption{Normalised observed duration distribution of the entire {\it Swift} GRB sample. The solid line corresponds to our best fit, while the black dashed (dotted) lines show the separate contributions of collapsar (non-collapsar) objects. The blue, vertical line marks the breakout time. The binning of the data and the corresponding poissonian error bars help visualisation, but have not been used for the fit (see Appendix \ref{sec:appendix} for details). We also show the assumed power-law distribution $p_{\rm e}$ (Eq. \ref{eq:Ne_parametric}), suitably rescaled by a factor $f_{\rm out}^{-1}$. The blue, dashed component corresponds to durations longer than $\hat{T}_{\rm b}$, i.e. to jets breaking out from the host star and powering the GRB prompt gamma-ray emission. The red, dotted component is extrapolated down to $T_{\rm min}=\left(1+\hat{z}\right)t_{\rm min}\sim 40\;$s and shows the case where all Type Ib/c SNe have a central engine launching a relativistic jet; for more details on this point we refer to Eq. \eqref{eq:tmin2} and the corresponding discussion.
}
\label{fig:distr}
\end{figure*}

\subsection{Definition of the model}

We denote the duration of the central engine jet-launching activity by $t_{\rm e}$, and the minimal $t_{\rm e}$ required for the jet to break out of the stellar envelope by $t_{\rm b}$. The energy ejected during the interval $0<t<t_{\rm b}$ is dissipated at the head of the jet while it propagates inside the progenitor star. The jet drives a shock into the stellar material that heats it up, pushing it aside and inflating a high-pressure cocoon that helps collimate the jet itself. If the central source does not remain active for a sufficiently long time, $t_{\rm e}<t_{\rm b}$, then most of the jet energy is dissipated at its head while still well inside its progenitor star. In this case, the jet is choked
and fails to make it out of the star, and its energy can only contribute to the associated supernova explosion. In the borderline case, when $t_{\rm e}$ is only slightly less than $t_{\rm b}$, the jet can put at most a modest amount of energy into a mildly relativistic outflow.

On the other hand, if the central engine is active for long enough, $t_{\rm e}>t_{\rm b}$, the jet can make it out of the progenitor star.
Once the jet's head breaks out of the stellar envelope it quickly accelerates and the jet material that is subsequently (i.e. at $t_{\rm b}<t<t_{\rm e}$) ejected can reach ultra-relativistic Lorentz factors and form a GRB, powering the observed prompt gamma-ray emission. The duration of the resulting prompt gamma-ray emission, $t_\gamma$, is expected to be similar to the post-breakout jet launching time, $t_\gamma=t_{\rm e}-t_{\rm b}$. Times denoted here with a lower-case $t$ are measured in the rest frame of the GRB central engine, while the observed GRB duration, $T_\gamma=(1+z)t_\gamma$ (where $z$ is the GRB's redshift), is longer by a factor $(1+z)$ that accounts for cosmological time dilation.

Therefore, following \cite{Bromberg2012}, we assume that the probability distribution of GRB durations, $p_\gamma$, is related to the probability distribution of intrinsic engine activity times $p_{\rm e}$ by 
\begin{equation}
\label{eq:time_distr}
p_\gamma\left(t_\gamma|t_{\rm b},z\right)\text{d}t_\gamma \propto p_{\rm e}\left(t_{\rm b}+t_\gamma|t_{\rm b},z\right)\text{d}t_\gamma\;,
\end{equation}
where both $p_\gamma$ and $p_{\rm e}$ can in principle depend on $t_{\rm b}$ and $z$.
As probability distributions they are normalised to unity:
\begin{equation}
\label{eq:norm}
\int_{0}^\infty p_\gamma\left(t_\gamma|t_{\rm b},z\right)\text{d}t_\gamma = \int_{0}^\infty p_{\rm e}\left(t_{\rm e}|t_{\rm b},z\right)\text{d}t_{\rm e} = 1\;.
\end{equation}
For a given $t_{\rm b}$ and $z$, the fraction $f_{\rm out}(t_{\rm b},z)$ of GRB jets launched by the central engine that make 
it out of the star is
\begin{equation}
\label{eq:fout1}
f_{\rm out}(t_{\rm b},z)=\int_{t_{\rm b}}^\infty p_{\rm e}\left(t_{\rm e}|t_{\rm b},z\right)\text{d}t_{\rm e}\;,
\end{equation}
which also provides the relative normalization for Eq.~(\ref{eq:time_distr}),
\begin{equation}
\label{eq:time_distr2}
p_\gamma\left(t_\gamma|t_{\rm b},z\right)\text{d}t_\gamma =
\frac{p_{\rm e}\left(t_{\rm b}+t_\gamma|t_{\rm b},z\right)\text{d}t_\gamma}{f_{\rm out}(t_{\rm b},z)}\;.
\end{equation}
Introducing the distribution of jet breakout times at a given redshift $p_{\rm b}(t_{\rm b}|z)$ we can write:
\begin{equation}
\label{eq:fout2}
f_{\rm out}(z)=\int_{0}^\infty p_{\rm b}\left(t_{\rm b}|z\right)f_{\rm out}(t_{\rm b},z)\text{d}t_{\rm b}\;.
\end{equation}
Using the GRB redshift distribution, $p_{\rm z}(z)$, we can finally obtain the total fraction of jets that make it out of the star:
\begin{eqnarray}
\label{eq:fout3}
f_{\rm out}&=&\int_{0}^\infty p_{\rm z}(z)\text{d}z f_{\rm out}(z),
\\ \nonumber
&=& \int_{0}^\infty p_{\rm z}(z)\text{d}z\int_{0}^\infty p_{\rm b}\left(t_{\rm b}|z\right)\text{d}t_{\rm b}\int_{t_{\rm b}}^\infty p_{\rm e}\left(t_{\rm e}|t_{\rm b},z\right)\text{d}t_{\rm e}\;.
\end{eqnarray}
The predicted long-soft GRB duration distribution is given by
\begin{eqnarray}
\label{eq:GRB1}
P_\gamma(T_\gamma) &=&  \int_{0}^\infty \frac{p_{\rm z}(z)\text{d}z}{1+z},
\int_{0}^\infty p_{\rm b}\left(t_{\rm b}|z\right)\text{d}t_{\rm b}p_\gamma\left(t_\gamma=\frac{T_\gamma}{1+z}|\, t_{\rm b},z\right)
\\ \nonumber
&=& \int_{0}^\infty \frac{p_{\rm z}(z)\text{d}z}{1+z}\int_{0}^\infty p_{\rm b}\left(t_{\rm b}|z\right)\text{d}t_{\rm b} 
\frac{p_{\rm e}\left(t_{\rm e}=t_{\rm b}+\frac{T_\gamma}{1+z}|\,t_{\rm b},z\right)}{f_{\rm out}(t_{\rm b},z)}\;.
\end{eqnarray}

We now make the following simplifying assumptions:
\begin{enumerate}
\item  The GRB redshift distribution is a delta function at a typical redshift $\hat{z}$,
\begin{equation} 
 p_{\rm z}\left(z\right) = \delta\left( z - \hat{z}\right)\;.
\end{equation}
\item The breakout time distribution is independent of redshift and is a delta function corresponding to a single breakout time $\hat{t}_{\rm b}$ in the rest frame of the star,
\begin{equation}\label{eq:p_b}
p_{\rm b}(t_{\rm b}|z) = \delta\left(t_{\rm b} - \hat{t}_{\rm b}\right)\;.
\end{equation}
\item  We ignore the possible dependence of $p_{\rm e}$ on $t_{\rm b}$ and $z$, i.e. 
\begin{equation}
p_{\rm e}\left(t_{\rm e}|t_{\rm b},z\right) = p_{\rm e}\left(t_{\rm e}\right)\;.
\end{equation}
\end{enumerate}

Assumption (i) follows \citet{Bromberg2012} and is motivated by the fact that the observed GRB redshift distribution, $p_{\rm z}\left(z\right)$, is rather involved and the selection function depends on the details of the detector such as its sensitivity, energy range and trigger algorithms, which vary between GRB samples from different instruments. As discussed in Section \ref{sec:caveats}, results are largely unchanged if one relaxes this assumption using the GRB subsample with an exact redshift determination.
Assumptions (ii) and (iii) are equivalent to neglecting any possible correlation between $\hat{t}_{\rm b}$ and $z$, and $t_{\rm e}$ and $t_{\rm b}$ respectively. In the absence of a better theoretical understanding of the underlying physics and/or an improved data statistics, we have found these to be the most reasonable assumptions.

Under these approximations,
\begin{equation}\label{eq:approx1a}
f_{\rm out}=\int_{\hat{t}_{\rm b}}^\infty p_{\rm e}\left(t_{\rm e}\right)\text{d}t_{\rm e}\;,
\end{equation}
\begin{equation}\label{eq:approx1b}
P_\gamma(T_\gamma) =  \frac{p_\gamma\left(t_\gamma=\frac{T_\gamma}{1+\hat{z}}\right)}{1+\hat{z}} =
\frac{p_{\rm e}\left(t_{\rm e}=\hat{t}_{\rm b}+\frac{T_\gamma}{1+\hat{z}}\right)}{(1+\hat{z})f_{\rm out}}\;.
\end{equation}
Note that for $T_\gamma\ll \left(1+\hat{z}\right)\hat{t}_{\rm b}$ we have $P_\gamma\left(T_\gamma\right)~\approx~p_{\rm e}\left(\hat{t}_{\rm b}\right)/(1~+~\hat{z})f_{\rm out}(\hat{t}_{\rm b},\hat{z})$, which is independent of $T_\gamma$; $P_\gamma$ therefore flattens according to observations \citep[see also][]{Bromberg2012}.

For $T_\gamma\gg \left(1+\hat{z}\right)\hat{t}_{\rm b}$ we have $p_{\rm e}\left(t_{\rm e}\right)\propto p_\gamma\left(t_\gamma=t_{\rm e}\right)$ so that the observations in this regime directly reflect the
functional form of $p_{\rm e}\left(t_{\rm e}\right)$.\footnote{More generally, Eq.~(\ref{eq:GRB1}) includes also the effects of  distributions in $z$ and $t_{\rm b}$ ($p_{\rm z}$ and $p_{\rm b}$ respectively). However, the typical fractional width of $1+z$ is $\sigma_{\rm 1+z}/\langle 1+z\rangle\sim 1$, so that a comparable or smaller width for $ p_{\rm b}\left(t_{\rm b}\right)$ would result mainly in a moderate smoothing of the break in $P_\gamma\left(T_\gamma\right)$ near $\hat{T}_{\rm b}$, but would not change the asymptotic power-law index $\alpha$. Since $\alpha$ is sampled through $P_\gamma\left(T_\gamma\right)$ over a finite range in $T_\gamma$, this can still have some effect on its inferred value of $\alpha$, but as discussed in Section \ref{sec:caveats} for $p_{\rm z}\left(z\right)$ this effect is not very large.}
Since the observed duration distribution of long-soft GRBs is consistent with a power law at the longest durations (well above the end of the plateau, which is identified with the breakout time; \citealt{Bromberg2012}), this suggest that the simplest possible $p_{\rm e}$ consistent with observations is a power law above some $t_{\rm min}$:
\begin{equation}\label{eq:Ne_parametric}
p_{\rm e}\left(t_{\rm e}\right) =
\begin{cases}
\frac{\alpha-1}{t_{\rm min}}
\left(\frac{t_{\rm e}}{t_{\rm min}}\right)^{-\alpha}& \qquad\text{for}\quad t_{\rm e} \geq t_{\rm min} \;, \\
0 &\qquad\text{for}\quad t_{\rm e}<t_{\rm min}\;.
\end{cases}
\end{equation}
Substitution of Eq.~\eqref{eq:Ne_parametric} into
Eq.~\eqref{eq:approx1a} gives
\begin{equation}
\label{eq:fout}
f_{\rm out} = \left(\frac{t_{\rm min}}{\hat{t}_{\rm b}}\right)^{\alpha-1}\;.
\end{equation}
Defining $\hat{T}_{\rm b} = (1+\hat{z})\hat{t}_{\rm b}$, using the relation $t_{\rm e}=t_{\rm b}+t_\gamma$,
and substituting Eqs.~\eqref{eq:Ne_parametric} and \eqref{eq:fout} into Eq.~\eqref{eq:approx1b}, we obtain
\begin{equation}
\label{eq:Pe1}
p_\gamma\left(t_\gamma\right)=\frac{\alpha-1}{\hat{t}_{\rm b}}
\left(1+\frac{t_\gamma}{\hat{t}_{\rm b}}\right)^{-\alpha}\;,\quad
P_\gamma\left(T_\gamma\right)=\frac{\alpha-1}{\hat{T}_{\rm b}}
\left(1+\frac{T_\gamma}{\hat{T}_{\rm b}}\right)^{-\alpha}\;.
\end{equation}
This functional form has the minimum number of parameters necessary to reproduce the properties of the observed duration distribution: (i) the power law index, $\alpha$; (ii) the plateau, which is associated to the properties of the progenitor star through $\hat{t}_{\rm b}$. In particular, note that $P_\gamma$ is independent of $t_{\rm min}$.

\subsection{Fitting the observed GRB time distribution}

From now on we consider the GRBs detected by {\it Swift},\footnote{\url{http://swift.gsfc.nasa.gov/archive/grb_table/}} and use a fiducial redshift of $\hat{z}=2$, which is the average redshift of the GRB sample we are considering. Figure \ref{fig:distr} shows the normalized observed duration distribution of the entire {\it Swift} GRB sample up to 16 May 2017, which contains 1130 GRBs. The observed GRB duration $T_\gamma$ is taken to be $T_{90}$, i.e. the interval over which the central 90\% of the photons from the GRB are detected. Note that since we are using a probability per unit time (i.e. duration), $P_\gamma(T_\gamma)\propto \text{d}N_{\rm GRB}/\text{d}T_\gamma$, this is different from the usual representation, $\text{d}N_{\rm GRB}/\text{d}\log(T_\gamma)~=~T_\gamma(\text{d}N_{\rm GRB}/\text{d}T_\gamma)$. The plateau predicted by the collapsar model extends for more than an order of magnitude, in the range $T_\gamma\sim 1\mhyphen20\text{ s}$.

It is well known that at short durations the GRB distribution is dominated by non-collapsar objects, which do not originate from the collapse of a massive star, but are generally believed to arise from the coalescence of a compact binary system (e.g. \citealt{Nakar2007}). Accounting for this different type or class of GRBs and separating their relative contribution from that associated to core-collapse GRBs is a delicate matter. We approach this problem by phenomenologically modelling the duration distribution of non-collapsars using a log-normal distribution (black dotted line in Figure \ref{fig:distr}).\footnote{Our conclusions rely on the values of $\alpha$ and $\hat{T}_{\rm b}$, which depend on the shape of the GRB time distribution at relatively long durations, $T_\gamma~\gtrsim~\hat{T}_{\rm b}~\sim~170\;$s. Since at these durations the non-collapsar contribution is expected to be negligible, we do not expect the results of this paper to change if one uses a functional form different than a lognormal to model the time distribution of non-collapsar GRBs.} We use a maximum-likelihood method to constrain parameters for both short and long GRBs simultaneously. Given the high dimensionality of the parameters space involved, we used a Markov-Chain Monte Carlo method to explore it. Details of this fitting method are described in Appendix \ref{sec:appendix}. Here we provide only the best-fit values (and the associated 1$\sigma$ confidence intervals) for the parameters of the collapsar-GRBs duration distribution (black, dashed line in Figure~\ref{fig:distr}):
\begin{equation}
\label{eq:fit_res}
\alpha=4.2_{-0.5}^{+0.6}\;, \qquad \hat{T}_{\rm b}=170_{-30}^{+40}\text{ s}\;.
\end{equation}
Figure \ref{fig:distr} also shows the corresponding central engine activity time power-law distribution, $p_{\rm e}$. Since here times are in the observer's frame, we have taken Eq.~\eqref{eq:Ne_parametric} with $t_{\rm e}\to\left(1+\hat{z}\right)t_{\rm e}$ and $t_{\rm min}\to\left(1+\hat{z}\right)t_{\rm min}$. We have furthermore rescaled $p_{\rm e}$ by a factor $f_{\rm out}^{-1}$ to show more clearly the correspondence with $p_\gamma$ in the regime $t_{\rm e}\gg \hat{t}_{\rm b}$.

Our fit implies a breakout time that is significantly longer than the end of the plateau in the GRB duration distribution (see the right, blue dashed vertical line in Figure~\ref{fig:distr}). Note that the end of the plateau was used as an estimate of $t_{\rm b}$ by previous works, but instead provides only a lower limit. The reason is that the GRB duration distribution is very steep at the longest durations (with $\alpha\sim 4.2$, following $p_{\rm e}\propto t_{\rm e}^{-\alpha}$), and our Eq. \eqref{eq:Pe1} predicts that the duration distribution is already suppressed relative to the plateau by a factor $2^{-\alpha}\sim 0.05$, i.e. by over an order of magnitude, at $T_\gamma=\hat{T}_{\rm b}$.

For our fiducial $\hat{z}\sim 2$, our results correspond to a breakout time in the rest frame of the star of $\hat{t}_{\rm b}\sim 60\text{ s}$. Theoretical predictions for cold, hydrodynamic jets generally yield lower values, of the order of $\hat{t}_{\rm b}\sim 10\mhyphen15\text{ s}$ \citep[e.g.][]{Bromberg2011,Bromberg2015}.
This tension indicates that some properties of the stellar structure may be different than what is usually assumed. We will return on this point in Section \ref{sec:dis}, where we show that the reason for such a large $t_{\rm b}$ could be a low-density, extended envelope surrounding the progenitors of long-soft GRBs.

\subsection{How robust are our results?} \label{sec:caveats}

\subsubsection{Redshift distribution, $p_{\rm z}\left(z\right)$}

We have performed some further tests to verify the robustness of our results. First, we test our approximation that $\hat{z}=2$ for all GRBs in the entire {\it Swift} sample, containing 1130 GRBs, which we have used in the fit shown in Figure~\ref{fig:distr}. To simplify the analysis, we have assumed that this is valid even for GRBs with a measured redshift. Do the results change by taking into account the GRBs redshift information? To test this, we have performed two additional analyses. In these we have:
{
\renewcommand{\theenumi}{
{\bf Analysis \Alph{enumi}}:}
\begin{enumerate}
 \item (i) selected the subsample of {\it Swift} GRBs with a known redshift (340 out of 1130 GRBs); (ii) derived the corresponding intrinsic duration distribution $p_\gamma\left(t_\gamma\right)$, where $t_\gamma~=~T_\gamma/\left(1+z\right)$ is measured in the rest frame of the star; (iii) fitted $p_\gamma$ to find $\hat{t}_{\rm b}$ and $\alpha$. Selecting GRBs with a known redshift reduces the available sample by $70\%$ on average, and by $90\%$ for GRBs with $T_\gamma<1\text{ s}$, which typically have a lower luminosity. Since information on the duration distribution of the non-collapsars objects is almost completely lost, we simply use Eq.~\eqref{eq:Pe1} as a fitting formula (i.e. neglect the small non-collapsar contribution). It would be unreliable to constrain the three additional parameters related to the non-collapsar GRB duration distribution (i.e. $f_{\rm nc}$, $\mu$ and $\sigma$; see Appendix \ref{sec:appendix}) on the basis of $\sim$ ten events.
 We find $\alpha=3.7_{-0.6}^{+1.1}$ and $\hat{t}_{\rm b}=54_{-15}^{+29}\text{ s}$, consistent with the results found above using the entire {\it Swift} sample.
 
\item (i) taken the observed duration distribution, $P_\gamma\left(T_\gamma\right)$, of the entire {\it Swift} sample and derived the corresponding $p_\gamma\left(t_\gamma\right)$ using $t_\gamma= T_\gamma/\left(1+z\right)$. When the GRB's redshift $z$ is not known, we assign a probability distribution $p_{\rm z}\left(z\right)$ equal to the observed {\it Swift} GRB redshift distribution (i.e. a discrete distribution with equal probability for each of the 340 measured redshift values);
(ii) fit the resulting $p_\gamma\left(t_\gamma\right)$, which takes into account the redshift as implied by the de-convolution.  In this case we find $\alpha=3.5_{-0.3}^{+0.6}$ and $\hat{t}_{\rm b}=46_{-7}^{+17}\text{ s}$, which again is consistent with the previous results. This method has the advantage to exploit both (i) the entire GRB sample and (ii) the redshift information. However, all the possible correlations between $t_\gamma$ and $z$ are lost; this is particularly relevant for non-collapsars that are typically detected at lower redshifts.
\end{enumerate}
}
We finally note that, regardless of the GRB sample considered, (i) the best fit value for the breakout time, $\hat{t}_{\rm b}\gtrsim 45\text{ s}$, is a factor $\gtrsim 3\mhyphen 5$ larger than the usual result for compact progenitors, $t_{\rm b}~\sim~10~\mhyphen~15\text{ s}$; (ii) the duration distribution of the central engines is quite steep, consistent with a power law index $\alpha\gtrsim 3.5$. Moreover, all the result presented in the following remain consistent within uncertainties with the fiducial ones if one changes the GRB sample and/or the fitting method; hence, our conclusions are largely unaffected by this issue.

\subsubsection{Central engine activity time distribution, $p_{\rm e}\left(t_{\rm e}\right)$}

Second, we have examined our assumption regarding the functional form of $p_{\rm e}(t_{\rm e})$, Eq.~\eqref{eq:Ne_parametric}. Indeed, while at $t_{\rm e}\gg\hat{t}_{\rm b}$ Eq. \eqref{eq:Ne_parametric} is well constrained by observations, at $t_{\rm e}<\hat{t}_{\rm b}$ the functional form of the central engine activity time distribution does not affect the predicted GRB duration distribution, $P_\gamma\left(T_\gamma\right)$. In the absence of further information and in order to avoid introducing additional parameters, we found it most reasonable to simply extend the observationally constrained functional form also to the regime where $p_{\rm e}(t_{\rm e})$ cannot be directly probed.

In general, the observed flattening of   $P_\gamma\left(T_\gamma\right)$ at short $T_\gamma$ may be obtained by one of the following two reasons or by a combination of the two: (i) the imprint of the breakout time, $\hat{t}_{\rm b}$, which we have discussed above; (ii) an intrinsic flattening of $p_{\rm e}(t_{\rm e})$ at $t_{\rm e}\lesssim\hat{t}_{\rm e}$.
It is important to note that $\hat{t}_{\rm e}$ depends on the properties of the stellar core, which has a typical radius of $\sim 10^8\text{ cm}$, while $\hat{t}_{\rm b}$ depends on the surrounding envelope, which extends out to $\gtrsim 10^{11}\text{ cm}$. Since the properties of the core and of the envelope are expected to be weakly coupled (e.g. \citealt{Crowther2007}), the most reasonable {\it a priori} assumption is that these time scales are unrelated, i.e. $\hat{t}_{\rm e}\ll\hat{t}_{\rm b}$. One possibility, for example, is that $\hat{t}_{\rm e}\sim t_{\rm ff}\ll \hat{t}_{\rm b}$, where $t_{\rm ff}\sim 0.02\mhyphen 2\;$s is the free-fall time of the stellar iron core; this value of $t_{\rm ff}$ corresponds to core densities of $\sim 10^6\mhyphen10^{10}{\rm\;g\;cm^{-3}}$ (see for example \citealt{Janka2012,Burrows2013}).

We now discuss the possible deviations of $p_{\rm e}$ from a power law. A useful example is if $p_{\rm e}\!\left(t_{\rm e}\right)$ has the exact functional form as $p_\gamma\!\left(t_\gamma\right)$ in Eq.~\eqref{eq:Pe1} with $t_\gamma\to t_{\rm e}$ and $\hat{t}_{\rm b}\to\hat{t}_{\rm e}$. According to Eqs.~\eqref{eq:approx1a} and \eqref{eq:approx1b} this would imply a predicted duration distribution $p_\gamma\left(t_\gamma\right)$ and $P_\gamma\left(T_\gamma\right)$ with the exact functional for as in Eq.~\eqref{eq:Pe1} only with the substitution $\hat{t}_{\rm b}\to\hat{t}_{\rm b}+\hat{t}_{\rm e}$ and $\hat{T}_{\rm b}\to\hat{T}_{\rm b}+\hat{T}_{\rm e}$, respectively, where $\hat{T}_{\rm e}=(1+\hat{z})\hat{t}_{\rm e}$. Since it has the exact functional form that we have used in our fit, we know that it provides a good fit to the data. In this case our results imply that $\hat{t}_{\rm b}+\hat{t}_{\rm e}\sim60\;$s, which leads to one of the following options: (i) $\hat{t}_{\rm e}\ll\hat{t}_{\rm b}\sim 60\;$s, and $\hat{t}_{\rm e}$ plays a role very similar to $t_{\rm min}$ in our model, only with a more moderate break at the shortest $t_{\rm e}\sim\hat{t}_{\rm e}$; (ii) $\hat{t}_{\rm b}\ll \hat{t}_{\rm e}\sim60\;$s, in which case the break in $P_\gamma\left(T_\gamma\right)$ is primarily caused by an intrinsic break in $p_{\rm e}\!\left(t_{\rm e}\right)$.
However, a typical engine activity time of $\hat{t}_{\rm e}\sim60\;$s appears hard to achieve for long-soft GRB progenitor stars, and there is also no {\it a priori} reason why $p_{\rm e}\!\left(t_{\rm e}\right)$ should be flat below such a break; this would require an extra free parameter (the power-law index below the break) compared to the single power law that we have considered.
A third possibility is that (iii) $\hat{t}_{\rm e}\sim\hat{t}_{\rm b}\sim 30\;$s, which would require a fine tuning of presumably unrelated parameters, and therefore seems unlikely given our current understanding of the relevant physics.

One could also try to model $p_{\rm e}$ using a log-normal distribution, $p_{\rm e}\left(t_{\rm e}\right)\propto \exp\left[-\left(\ln\left(t_{\rm e}\right) -\mu\right)^2/2\sigma^2\right]/t_{\rm e}$, where $\mu \equiv \ln\left(\hat{t}_{\rm e}\right)$. Of course, this choice for $p_{\rm e}$ involves one additional parameter, which is hard to constrain from the data. One can therefore, e.g., fix $\hat{t}_{\rm e}$ before making the fit in order to reduce the number of free parameters. Assuming that $\hat{t}_{\rm e}\sim t_{\rm ff}\lesssim 2\;$s, $p_{\rm e}$ indeed deviates by $\lesssim 30\%$ from a power law when $t_{\rm e}>\hat{t}_{\rm b}$ (within the observed range of GRB durations $t_{\rm e}\sim t_\gamma\lesssim 300\;$s), and the results of the fit are not significantly affected.

In other words, taking a power law for $p_{\rm e}\!\left(t_{\rm e}\right)$ means assuming that it does not have an associated timescale $\hat{t}_{\rm e}$ near $\hat{t}_{\rm b}$, which would require undesirable fine-tuning. Avoiding such a fine-tuning is ultimately equivalent to assuming that $p_{\rm e}$ follows Eq. \eqref{eq:Ne_parametric} with $t_{\rm min}\ll\hat{t}_{\rm b}$.

\section{Constraints on the central engine}
\label{sec:constraints}

\subsection{Fraction of Type Ib/c SNe with a long-lived, anisotropic energy injection}

We are particularly interested in constraining the fraction $f_{\rm jet}$ of SNe Ib/c that possess a central engine launching a jet. For this purpose we shall compare:
\begin{enumerate}
\item the observed rate of GRBs of the long-soft class\footnote{Here we refer to GRBs that originate from the collapse of a massive star, which are known to be associated with SNe Ib/c.} at the typical redshift $\hat{z}=2$, $R_{\rm GRB}(\hat{z})$;
\item the observed rate of SNe Ib/c at the same typical redshift, $R_{\rm Ibc}(\hat{z})$.
\end{enumerate}
The ratio of these rates provides an estimate of the fraction of CC SNe with a central engine launching a jet that (i) manages to break out of the star and (ii) points in the direction of the Earth.
 
The rate of CC SNe at the relevant $\hat{z}\sim 2$ is $R_{\rm CC}\left(\hat{z}\right)~\sim~3\times~10^5\text{ Gpc$^{-3}$ yr$^{-1}$}$ (\citealt{Strolger2015}; see also \citealt{Dahlen2004, Graur2011, Melinder2012, Dahlen2012, Cappellaro2015, Petrushevska2016}).
The fraction of CC SNe that belong to the Ib/c class is found to be $f_{\rm Ibc}\sim 0.4$ at a mean redshift $z\sim 0.25$ by \citet{Cappellaro2015}; combining with local estimates (e.g. \citealt{BoissierPrantzos2009,Li2011}), they concluded there is currently no evidence for $f_{\rm Ibc}$ evolving with redshift. Hence, we estimate the rate of SNe Ib/c at the redshift at the typical redshift $z\sim 2$ to be:
\begin{equation}
\label{eq:R_Ibc}
R_{\rm Ibc}\left(\hat{z}\right)=f_{\rm Ibc}R_{\rm CC}\left(\hat{z}\right)\sim 1.2\times 10^5\text{ Gpc$^{-3}$ yr$^{-1}$}\;.
\end{equation}
Unfortunately, there seems to be no measurements of $f_{\rm Ibc}$ at higher redshifts at the time of writing. However, even in the extreme case of $f_{\rm Ibc}=1$, our main conclusions do not change significantly; we refer for more details to Appendix \ref{sec:typeII}, where we discuss the possible extension to the entire family of CC SNe.

To calculate the estimated rate of GRBs we have to account for (i) the fraction $f_{\rm jet}$ of SNe Ib/c that do have a jet; (ii) the fraction $f_{\rm out}$ of jets that make it out of the star; (iii) the beaming factor, $f_{\rm b}$, to find the fraction of jets pointing in the general direction of the Earth that are hence observable as GRBs (at least out to some redshift that depends on their luminosity and the detector's sensitivity). We assume a typical beaming factor of $f_{\rm b}\sim 10^{-2}$, corresponding to an opening half-angle $\theta_{\rm jet}\sim 8^\circ$ for a double-sided jet (e.g. \citealt{Frail2001,Bloom2003,Guetta2005,Friedman2005,LeDermer2007}). Using Eq.~\eqref{eq:R_Ibc}, we finally obtain
\begin{equation}
\label{eq:rate}
R_{\rm GRB}(\hat{z})=f_{\rm jet}f_{\rm out}f_{\rm b}R_{\rm Ibc}(\hat{z})\sim 1.2\times 10^3 f_{\rm jet} f_{\rm out} f_{\rm b,-2}\text{ Gpc$^{-3}$ yr$^{-1}$}\;,
\end{equation}
where $f_{\rm b,-2}=f_{\rm b}/10^2$.

The observed rate of long GRBs as a function of redshift is well fitted by a broken power law; the corresponding rate at the relevant $\hat{z}=2$ is given by \citet{WandermanPiran2010}\footnote{\citet{WandermanPiran2010} considered a sample of bursts with $L~>~10^{50}\text{ erg s$^{-1}$}$ and $t_\gamma>2\text{ s}$. To infer the rate of collapsar GRBs one should consider (i) the fraction of contaminating non collapsar objects ($\sim 20\%$); (ii) the fraction of missing collapsar GRBs, those shorter than $2\text{ s}$ ($\sim 40\%$; see \citealt{Bromberg2013}). The two effects tend to compensate, and the correction is well within uncertainties.} as
\begin{equation} \label{eq:ratedirect}
R_{\rm GRB}(\hat{z})\sim 10\text{ Gpc$^{-3}$ yr$^{-1}$}
\end{equation}
Comparing Eq. \eqref{eq:rate} with \eqref{eq:ratedirect} we can constrain the unknown parameters:
\begin{equation}
\label{eq:fout_1}
f_{\rm jet} f_{\rm out}\sim 8\times 10^{-3}f_{\rm b,-2}^{-1}\;.
\end{equation}
This estimate for $f_{\rm jet} f_{\rm out}$ depends on the redshift through the ratio $f_{\rm Ibc}R_{\rm CC}\left(z\right)/R_{\rm GRB}\left(z\right)$. Current observations are consistent with a common redshift evolution of $R_{\rm CC}\left(z\right)$ and $R_{\rm GRB}\left(z\right)$ at $z\lesssim 4$ (e.g. \citealt{RobertsonEllis2012}), where most of the GRBs are observed. In particular, note that the ratio of the beaming-corrected GRB volumetric rate to the CC SN volumetric rate is comparable with the local one (for $z\sim 0$ rates see \citealt{GuettaDellavalle2007}). Hence, any possible redshift dependence would come from the factor $f_{\rm Ibc}$; however, as discussed above, this factor has a small impact on our conclusions.

Since $f_{\rm out}\leq 1$, Eq. \eqref{eq:fout_1} strictly implies that
\begin{equation}
\label{eq:fjet_constraints}
0.01~\lesssim~f_{\rm jet}~\lesssim~1\;.
\end{equation}
Note that the constraints in Eq. \eqref{eq:fjet_constraints} are independent of any assumption on $p_{\rm e}$ and only rely on considerations of general character.

\subsection{Minimal activity time of the central engine}

In our phenomenological model, the lowest limit $f_{\rm jet}\sim 0.01$ would correspond to the extreme case when the distribution of the central engine activity times, $p_{\rm e}\propto t_{\rm e}^{-\alpha}$, is truncated at $t_{\rm min}=\hat{t}_{\rm b}$, which as noted above corresponds to an undesirable fine-tuning. It is therefore natural to ask where $p_{\rm e}$ should be truncated for different values of $f_{\rm jet}$.

At times larger than $\hat{t}_{\rm b}$, the time distribution of the central engines is consistent with a power-law profile, with a very steep index $\alpha\sim 4.2$. Extending the range of validity of this approximation down to $t_{\rm min}$ and combining Eq. \eqref{eq:fout} with Eq. \eqref{eq:fout_1}, we find
\begin{equation}
\label{eq:tmin}
f_{\rm jet} = 8\times 10^{-3}\left(\frac{r}{f_{\rm b,-2}}\right) \left(\frac{\hat{t}_{\rm b}}{t_{\rm min}}\right)^{\alpha-1}\;,
\end{equation}
where the spread of $r\equiv R_{\rm CC}R_{\rm GRB}/3\times 10^6\text{ Gpc$^{-6}$ yr$^{-2}$}$ parameterises our uncertainty on the rates.
At $z\sim 2$ we have $\delta\log_{10}\left(R_{\rm CC}\right)~\sim~0.3$ \citep{Strolger2015} and $\delta\log_{10}\left(R_{\rm GRB}\right)\sim 0.3$ \citep{WandermanPiran2010}.
We also consider the additional uncertainty due to the beaming factor, $\delta\log_{10}\left(f_{\rm b,-2}\right)\sim 0.4$ \citep{Liang2008,Racusin2009,Goldstein2016}. Hence, we end up with $\delta\log_{10}\left(r/f_{\rm b,-2}\right)\sim 0.6$.
Solving Eq. \eqref{eq:tmin} for $t_{\rm min}/\hat{t}_{\rm b}$, one can easily see that
\begin{equation}
\label{eq:tmin2}
\frac{t_{\rm min}}{\hat{t}_{\rm b}} = \left(\frac{0.008\,r}{f_{\rm jet}f_{\rm b,-2}}\right)^{\frac{1}{\alpha-1}}
= 0.22_{-0.08}^{+0.14}\times f_{\rm jet}^{-1/(\alpha-1)}\;,
\end{equation}
where the numerical value is for $\alpha=4.2_{-0.5}^{+0.6}$ and $\delta\log_{10}\left(r/f_{\rm b,-2}\right)=0.6$. Since $f_{\rm jet}\leq1$, this implies $t_{\rm min}\gtrsim13\;$s, which is an interesting result. The value of $t_{\rm min}$ corresponding to $f_{\rm jet}=1$, $t_{\rm min}\sim 13\;$s (accounting for cosmological time dilation, $T_{\rm min}=(1+\hat{z})t_{\rm min}$, with $\hat{z}=2$), is shown by the left, red vertical line in Figure~\ref{fig:distr}.

Hence, if $p_{\rm e}$ extends by a relatively small factor of $\sim 5$ below $\hat{t}_{\rm b}$ (i.e. down to $t_{\rm min}\sim 0.2\times\hat{t}_{\rm b}$), the number of central engines launching a jet would be comparable to that of all Type Ib/c SNe. Moreover, even a moderate $f_{\rm jet}=0.1$ requires an increasing fine tuning of the parameters, $t_{\rm min}/\hat{t}_{\rm b}=0.45_{-0.16}^{+0.26}$; according to considerations on the relative rates discussed in Section \ref{sec:concl}, this would correspond to the regime in which only {\it ll}GRBs and long GRBs have central engines launching a jet.

The collapsar model for long-soft GRBs may therefore be consistent with $f_{\rm jet}\sim 1$ (and therefore $f_{\rm out}\sim10^{-2}$), i.e. with most SNe Ib/c having a jet-launching central engine. Such a scenario is also supported by (i) the increasing number of transition objects detected between regular Type Ib/c SNe and long-soft GRBs, including {\it ll}GRBs and relativistic SNe (e.g. \citealt{Margutti2014}); (ii) the energy distribution of the ejecta of CC SNe \citep{Piran2017}; (iii) the morphology of CC SN remnants \citep{Bear2017}. We will discuss in more detail the implications of this possibility in Section \ref{sec:implications}.

\section{Discussion}
\label{sec:discussion}

\subsection{Implications for Type Ib/c SNe explosions}
\label{sec:implications}

According to the interpretation adopted here for the duration distribution of GRBs, the vast majority of central engines launching relativistic jets are active over times $t_{\rm e}<t_{\rm b}$ and their jets do not break out from the stellar envelope. However, these choked jets could have an important impact on the properties of the Type Ib/c SNe explosions (e.g. their energetics, asphericity, nucleosynthesis, light curves). Moreover, if $f_{\rm jet}\sim 1$, such an impact would be relevant for all Type Ib/c SNe. We now explore the implications of this possibility.

The total amount of energy released in the jets of long GRBs spans $\sim 3$ orders of magnitude, $E_{\rm jet}\sim 10^{49}\mhyphen10^{52}\text{ erg}$ (e.g. \citealt{KumarZhang2015} and references therein). Interestingly, this is comparable with both (i) the typical energy of CC SNe ejecta, $E_{\rm SN}\sim 10^{50}\mhyphen10^{52}\text{ erg}$ (e.g. \citealt{Burrows2013}); (ii) the binding energy of the envelope around the iron core in pre-SN stars, $U_{\rm b}~\sim~0.1~\mhyphen~2.5~\times~10^{51}\text{ erg}$ (depending on the initial mass and the metallicity; e.g. \citealt{Woosley2002}).\footnote{An additional issue arises from the fact that $E_{\rm jet}$ is measured for engines with typical durations $t_{\rm e}\gtrsim t_{\rm b}$, while here we are mainly interested in jets that do not break out of the star.
To derive the typical energy of these jets, one could extrapolate the positive correlations between the duration and the luminosity/energy of the GRBs. However, these correlations have a large scatter ($\sim 2$ orders of magnitude), and the most relevant duration/energy correlation is less clear (see for example \citealt{Hou2013}). Moreover, one should also consider the beaming (if known) and exclude non-collapsar objects, introducing additional uncertainties and reducing the available sample further. Hence, given such uncertainties, in the following we are not trying to be more quantitative on this point.}

Generally, the jet's energy is channeled into the cocoon that can also contribute to the SN explosion energy while its head is still inside the star (at $t<t_{\rm b}$), and into the relativistic ejecta that can power the GRB once they break out of the star (at $t>t_{\rm b}$).
If the jet fails to break out ($t_{\rm e}<t_{\rm b}$) then most of its energy is deposited into the stellar envelope, and can more effectively help to unbind it, contributing to the kinetic energy of the SN explosion.
The jet's contribution towards the SN explosion energy is therefore
\begin{equation}
\label{eq:deltaE}
\Delta E_{\rm SN,jet}\sim L_{\rm jet}\min(t_{\rm e},t_{\rm b}) \approx E_{\rm jet}\min(1,t_{\rm b}/t_{\rm e})\;.
\end{equation}
If $L_{\rm jet}t_{\rm e}$ increases with $t_{\rm e}$, the longest-living engines (corresponding to $t_{\rm e}\sim t_{\rm b}$) likely inject more energy into the SN explosion compared to those with $t_{\rm e}\ll t_{\rm b}$. This is consistent with the fact that SNe associated with GRBs are more energetic than average. In fact, as long as $t_{\rm e}\lesssim 2\times t_{\rm b}$, the majority of the jet's energy goes into the cocoon and/or towards the SN explosion. Since only $10\%$ of the long-soft GRBs lasts for $T_\gamma~>~T_{\rm b}\sim 170\text{ s}$, this is the most common case also for successful GRBs.

This contribution to the SN explosion (i.e. $\Delta E_{\rm SN,jet}$) could make the difference between a successful and a failed explosion, or even become the dominant channel. In this case, one would need $\Delta E_{\rm SN,jet}~\sim~U_{\rm b}+E_{\rm SN}$, a possibility that cannot be excluded {\it a priori} (the required $\Delta E_{\rm SN,jet}$ is lower if the energy is shared with only some fraction of the envelope in a strongly asymmetric SN).

The fact that a jet which fails to break out can still contribute to the explosion of a CC SN is particularly interesting given that the most popular explosion mechanism faces several difficulties.
In this scenario, neutrinos coming from the hot, inner core are absorbed by the outer layers of the star; neutrino heating establishes a sufficient pressure gradient that is sufficient to push part of the envelope outwards, eventually driving the observed explosions \citep{BetheWilson1985}. However, this requires a fraction $\sim 0.1\%$ of the neutrino energy to be reabsorbed by the outer layers of the star, and it is not completely understood how this can be achieved in practice. Though simulations of this process have made significant progress, the possibility for neutrinos to drive the explosion of CC SNe (and, in particular, of the most energetic ones) is still controversial (for recent reviews see \citealt{Janka2012,Janka2016}).

Hence, different possible contributions to the amount of energy required deserve careful consideration. An alternative scenario involves jet production via a magnetorotational mechanism during core collapse (\citealt{LeblancWilson1970}; see also \citealt{OstrikerGunn1971, Bisnovatyi-Kogan1971}). Although this is unlikely to be the dominant mechanism in all CC SNe (as discussed in Appendix \ref{sec:typeII}), our results suggest that it may be more common than previously thought.

Different authors (e.g. \citealt{Khokhlov1999, MaedaNomoto2003, Couch2009}) explored the possibility for CC SNe to be jet-driven. However, all these works focused on Newtonian jets, while a GRB-like central engine would launch a relativistic jet, which may have different effects on the SN explosion. An important attempt to unify the zoo of SNe explosions driven by relativistic jets was carried out by \citet{Lazzati2012}. These authors explored a wide range of durations of the central engine activity and two compact (with radius $(4.1\mhyphen4.8)\times 10^{10}\text{ cm}$) candidate stellar progenitors, while the total energy release was in the range $(0.3\mhyphen1.0)\times 10^{52}\text{ erg}$. They found that even if the jets are narrowly collimated (they used $\theta_{\rm jet}=10\degree$), their interaction with the star unbinds the stellar envelope, producing a stellar explosion.
In their simulations, the outcome of the explosion had a strong dependence on the duration of the engine activity, and they identified three regimes, based on the velocity of the ejecta. Only the longest-lasting engines (corresponding to $t_{\rm e}\gtrsim t_{\rm b}$ in the notation adopted here) were associated to successful GRBs. Engines with intermediate durations ($0.5\times t_{\rm b}\lesssim t_{\rm e}\lesssim t_{\rm b}$) produced relativistic SNe/{\it ll}GRBs, with a lower collimation with respect to long GRBs. Finally, they concluded that the engines with the shortest durations ($t_{\rm e}\ll t_{\rm b}$), {\it if they exist in nature}, result in stellar explosions that are dynamically indistinguishable from ordinary Type Ib/c SNe.

Of course, even if the jet alone manages to reproduce the dynamics of the SN explosion, it is still not the end of the story. In order to be successful, a SN model should also explain other features (i.e. the already mentioned asphericity, nucleosynthesis, light curves). Moreover, these features may vary with (i) the jet's properties, i.e. injection timescale $t_{\rm e}$, power $L_{\rm jet}$, opening angle $\theta_{\rm jet}$, composition (in particular thermal energy and degree of magnetisation); (ii) the envelope's properties (essentially mass and radius), which reflect in the breakout time $t_{\rm b}$. Further investigation is certainly required to understand if different combinations of these parameters can reproduce at least part of the wide variety of the observed events. Finally, note that the presence of a central engine launching a jet does not exclude any contribution from other channels, e.g. neutrinos.

\subsubsection{The geometry of the explosion}

In the most extreme but possible scenario in which all pre-SN stars produce bipolar jets similar to those powering GRBs, one would expect an intrinsically asymmetric engine working in all CC SNe. Due to the interaction with the stellar envelope, jets choked long before the breakout would produce more spherical blast waves. Hence, the asymmetry of the CC SN explosion should increase as $t_{\rm e}/t_{\rm b}$ approaches unity or, in the few cases when a GRB is produced, exceeds unity. Note that the argument below is still valid, or even reinforced, if the jet's energy is not dynamically relevant for the explosions of Type II SNe.

The observations of CC SNe, including both Type Ib/c and Type II, show abundant evidence of deviations from spherical symmetry. \citet{Wang2001} first noted a general trend for the asymmetry to increase with decreasing envelope mass and with increasing depth within the ejecta. Different authors (e.g. \citealt{Maeda2008, Modjaz2008, WangWheeler2008, Taubenberger2009,Cano2016}) have attempted a more comprehensive analysis, showing convincing evidence that all CC SNe from stripped-envelope stars are at least mildly non-spherical. This asymmetry is generally more accentuated than in Type II SNe, while Type Ic SNe accompanied by GRBs exhibit the highest degree of asymmetry. Hence, the observations of CC SN ejecta and their association with long-soft GRBs seem to be globally consistent with a jet-driven scenario.

\subsection{A common progenitor for {\it ll}GRBs and long-soft GRBs?}
\label{sec:dis}

In any attempt to identify common features among jet-driven SNe explosions, {\it ll}GRBs play a fundamental role as intermediate events between regular SNe (which we argue may hide a GRB-like jet) and long GRBs. Since due to their low luminosity they are more difficult to observe than regular GRBs, only five {\it ll}GRBs have been clearly detected to date.\footnote{These are GRBs with luminosity $\lesssim 10^{48}\text{ erg s$^{-1}$}$ for which a spectroscopically associated SN was observed -- {\it ll}GRB/SN: 980425/1998bw, 031203/2003lw, 060218/2006aj, 100316D/2010bh -- and the GRB 020903, which has a photometrically associated SN.} Different properties of {\it ll}GRBs with respect to long GRBs suggest a different emission mechanism. Specifically, {\it ll}GRBs may arise from jets that do not manage to break out from the star, thus failing to power the prompt emission as in long GRBs. However, if such jets are choked close enough to the surface (i.e. $t_{\rm e}\lesssim t_{\rm b}$), they can still produce a powerful shock breakout.

\citet{NakarSari2012} showed that the properties of relativistic shock breakouts are indeed in good agreement with the main observational signatures (typical energy, duration and peak photon energy) of {\it ll}GRBs. According to this interpretation, the breakout radius should be $R_{\rm ext}\approx 10^{13}\text{ cm}$, two orders of magnitude larger than the typical size of the compact progenitors (i.e. Wolf-Rayet stars; for a review see \citealt{Crowther2007}). This discrepancy can be explained if the star suffers strong mass losses prior to explosion, and is therefore surrounded by a low-density, extended envelope during the final stages of its evolution \citep{Margutti2015, Nakar2015}. Observations of the the SN 2006aj, associated with the {\it ll}GRB 060218, support this scenario: the light curve of the SN shows an early peak, which is likely due to the radiative cooling of an extended envelope with $R_{\rm ext}\sim 10^{13}\text{ cm}$ and $M_{\rm ext}\sim 0.01\;M_\odot$ \citep{Nakar2015}.

If there is an extended, low-density mass shell surrounding the star, the minimum engine working time needed to drive a successful GRB is significantly longer than the usual result for compact progenitors, which is $t_{\rm b}\sim 10-15\text{ s}$ (e.g. \citealt{Bromberg2011,Bromberg2015}). As discussed in Appendix \ref{sec:appendix2}, for a hydrodynamic jet propagating through an envelope with a flat density profile one finds
\begin{equation}
\label{eq:tb_ext}
t_{\rm b}\sim 67\left(\frac{L_{\rm iso}}{10^{51}\text{ erg s$^{-1}$}}\right)^{-1/2} \left(\frac{R_{\rm ext}}{10^{13}\text{ cm}}\right)^{1/2} \left(\frac{M_{\rm ext}}{10^{-2}M_\odot}\right)^{1/2}\text{ s}\;,
\end{equation}
where $R_{\rm ext}$ ($M_{\rm ext}$) is the radius (mass) of the extended envelope, $L_{\rm iso}\equiv L_{\rm jet}\theta_{\rm jet}^2/2$ is the jet's isotropic luminosity, which is taken as constant, and $\theta_{\rm jet}$ is the jet's half-opening angle. Note that $t_{\rm b}$ is shorter than the light crossing time of the star, which is the proper case when the head of the jet is relativistic.

Since the jet is more easily choked if there is a low-density, extended envelope surrounding the star, \citet{Nakar2015} suggested that the absence of such an envelope could make the difference between  a {\it ll}GRB and a long GRB. According to this interpretation, the duration $t_{\rm e}$ of the central engine is similar for both {\it ll}GRB and long GRBs, but the envelope is not present in the latter case. Hence, the breakout time for long GRBs would be $t_{\rm b}\sim 10\mhyphen15\text{ s}$, significantly shorter than the result of Eq. \eqref{eq:tb_ext}, and the jet manages more easily to break out from the star.

Our results suggest a different possible interpretation: fitting the duration distribution of long GRBs we find $t_{\rm b}\sim 60\text{ s}$, which is a factor $\sim 4\mhyphen 6$ larger than the usual result for compact progenitors (which would correspond to $t_{\rm b}\sim 10\mhyphen15\text{ s}$ in the rest frame of the star, with a weak dependence on the stellar parameters; e.g. \citealt{Bromberg2011}). Interestingly, given the uncertainties on all the parameters in Eq.~\eqref{eq:tb_ext}, a breakout time of $t_{\rm b}\sim 60\text{ s}$ is consistent with that inferred from the properties of {\it ll}GRBs progenitors. Hence, the progenitors of both long-soft and {\it ll}GRBs may be surrounded by low-mass, extended envelopes. According to this interpretation, the duration of the central engine, $t_{\rm e}$, makes the main difference between {\it ll}GRBs (corresponding to $t_{\rm e}<t_{\rm b}$) and long GRBs (corresponding to $t_{\rm e}>t_{\rm b}$), while $t_{\rm b}$ is similar for both the classes. Finally, note that the breakout time depends only on the product $M_{\rm ext}\,R_{\rm ext}$; hence, different combinations of the envelope's mass and radius can still result in the same $t_{\rm b}$.

\section{Conclusions}
\label{sec:concl}

\begin{figure*}{\vspace{3mm}} 
\centering
\includegraphics[width=0.99\textwidth]{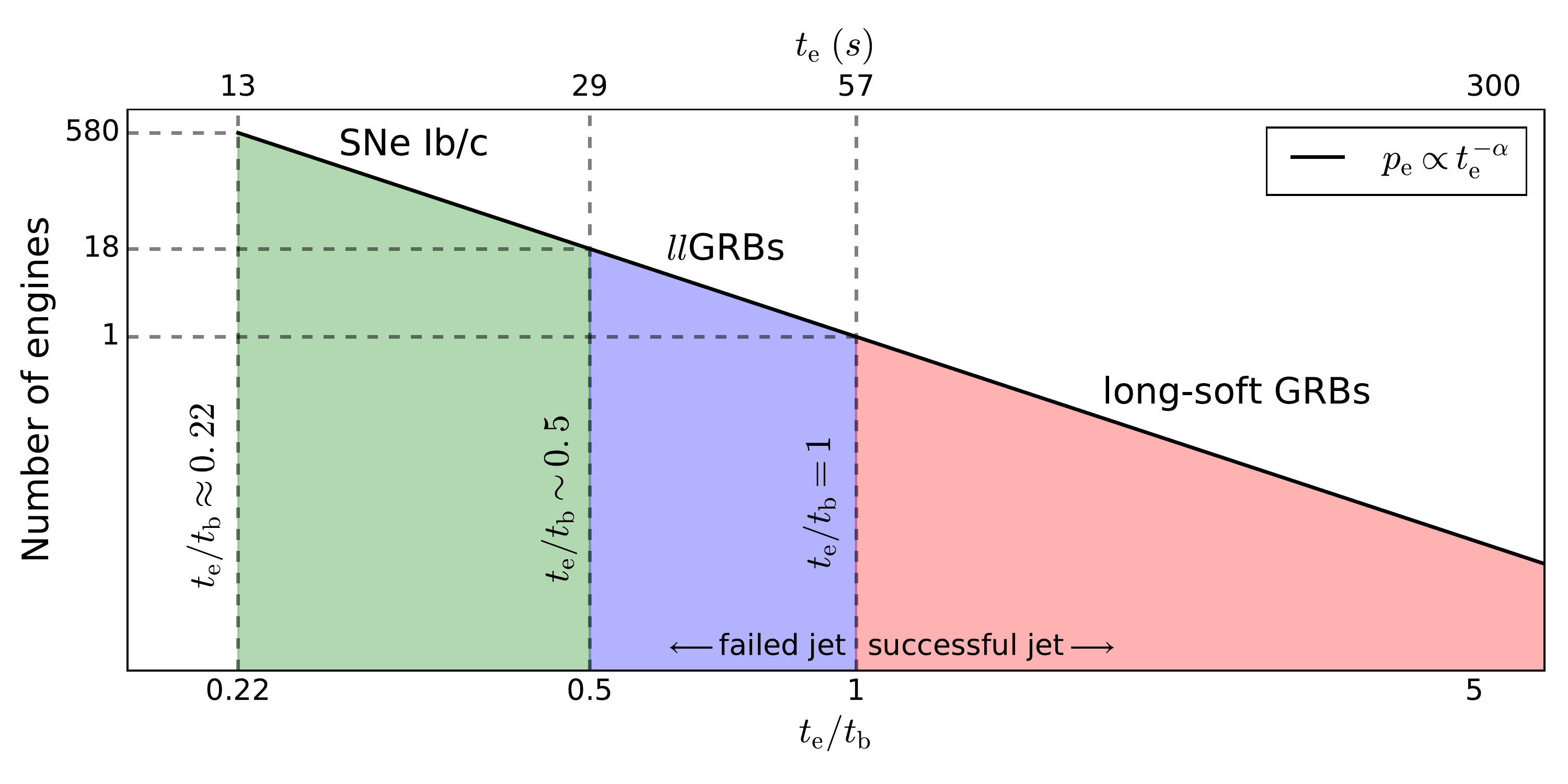}
\caption{Sketch of the outlined physical picture. Times are given in the proper frame of the star, and we use a logarithmic scale on both axes. The time distribution of the central engines is a steep power law, $p_{\rm e}\propto t_{\rm e}^{-\alpha}$ with $\alpha\sim 4.2$. Jets launched by the longest-lasting engines ($t_{\rm e}>t_{\rm b}$) break out of the star and power the prompt gamma-ray emission of long-soft GRBs, lasting for $t_\gamma=t_{\rm e}-t_{\rm b}$. Engines with intermediate durations ($0.5\lesssim t_{\rm e}/t_{\rm b}<1$, i.e. the jet is choked close to the surface) do not produce any prompt emission, while they correspond to powerful shock breakouts which can reproduce the properties of {\it ll}GRBs. If $p_{\rm e}$ is extrapolated down to shorter times, all the other jets ($0.2\lesssim t_{\rm e}/t_{\rm b}\lesssim 0.5$) deposit their energy deep into the star, and may significantly contribute to the explosion of regular SNe Ib/c. Extending $p_{\rm e}$ down to $t_{\rm e}/t_{\rm b}\sim 0.5$, one would reproduce the rate of broad-lined SNe, which make up a fraction $\lesssim 10\%$ of all Type Ib/c (e.g. \citealt{GuettaDellavalle2007, Drout2011}). The fact that broad-lined SNe could correspond to choked jets was indeed proposed by \citet{Modjaz2016}.}
\label{fig:main}
\end{figure*}

In the framework of the collapsar scenario, we have explored the possibility that SNe that do not produce a long-soft GRB also possess a jet which is choked inside the star. Our conclusions are summarised below.

\subsection{Summary of the phenomenological model}

We have found that the duration distribution of long-soft GRBs can be reproduced using a simple phenomenological model starting from two minimal assumptions: (i) there is a single jet-breakout time valid for all SNe, $t_{\rm b}=\hat{t}_{\rm b}$ (Eq.~\ref{eq:p_b}); (ii) the probability distribution of the central engine activity time, $t_{\rm e}$, is given by Eq.~\eqref{eq:Ne_parametric}:
\begin{equation}
p_{\rm e}\left(t_{\rm e}\right)\propto t_{\rm e}^{-\alpha} \qquad\text{for}\qquad t_{\rm e}>t_{\rm min}\nonumber\;,
\end{equation}
where $t_{\rm min}<t_{\rm b}$ and $\alpha$ are parameters of the model. This assumption is motivated by the shape of the observed GRB duration distribution at durations longer than $t_{\rm b}$ and by avoiding the introduction of an additional parameter whose value, to result in a distribution that qualitatively differs from the power law assumption, would require an undesirable and arbitrary fine-tuning, i.e. would require the breakout time and the typical central engine activity time, which are presumably unrelated, to be comparable.

Using the relation $t_\gamma=t_{\rm e}-t_{\rm b}$, where $t_\gamma$ is the GRB duration in the proper frame of the star, we find that these assumptions can indeed provide a good fit to the observed GRB duration distribution. Eq. \eqref{eq:fit_res} gives our best-fit values
\begin{equation}
\label{eq:final}
\hat{t}_{\rm b}=57_{-10}^{+13}\text{ s} \qquad\qquad \alpha=4.2_{-0.5}^{+0.6}\nonumber\;,
\end{equation}
where $\hat{t}_{\rm b}$ is measured in the proper frame of the star. We expect our results to be fairly robust against selection effects (see the discussion in Appendix \ref{sec:appendix}). Below we summarise the potential implications of these results for {\it ll}GRBs, long-soft GRBs and Type Ib/c SNe.

\subsection{Implications}

\subsubsection*{Gamma Ray Burst progenitors}

In the collapsar scenario, long-soft GRBs correspond to jets with $t_{\rm e}>t_{\rm b}$ that manage to break out of the star and power the prompt gamma-ray emission at times $t_{\rm b}<t<t_{\rm e}$. Instead, jets that are ``choked'' close enough to the surface (i.e. those in the regime $t_{\rm e}\lesssim t_{\rm b}$) do not produce any prompt gamma-ray emission. However, these jets still produce a powerful, quasi-spherical shock breakout which can be responsible for the observed emission of {\it ll}GRBs.

Modelling their properties as shock breakouts, \citet{NakarSari2012} suggested {\it ll}GRBs to be surrounded by an envelope which is two orders of magnitude more extended than the typical size of Wolf-Rayet stars, and is possibly due to strong mass losses prior to explosion. This result was further confirmed by modelling the early light curve of the SN 2006aj (associated to the {\it ll}GRB 060218), in which case one finds $R_{\rm ext}\sim 3\times 10^{13}\text{ cm}$ ($M_{\rm ext}\sim 0.01\;M_\odot$) for the radius (mass) of the envelope \citep{Nakar2015}.

Our inferred breakout time for long-soft GRBs ($\hat{t}_{\rm b}\sim60\text{ s}$ in the proper frame of the star) is unusually large for hydrogen-stripped progenitors, and suggests the presence of an extended, low-density envelope surrounding the pre-SN star. Hence, both {\it ll}GRBs and long-soft GRBs may be consistent with similar envelope properties.

Despite large uncertainties, the volumetric rate of {\it ll}GRBs appears to be a factor $\approx 10$ larger than for long GRBs (e.g. \citealt{Soderberg2006, GuettaDellavalle2007}). Due to the steepness of the central engine time distribution, using a power-law model ($p_{\rm e}\propto t_{\rm e}^{-\alpha}$ with $\alpha\sim 4.2$) such an increase by a factor of ten can be achieved if {\it ll}GRBs are produced from a relatively narrow range of engine activity, namely $0.5<t_{\rm e}/t_{\rm b}<1$. Interestingly, in their simulations, \citet{Lazzati2012} found these intermediate class explosions to occur for $0.6<t_{\rm e}/t_{\rm b}<1$ or $0.45<~t_{\rm e}/t_{\rm b}<1$, depending on the total energy released in the jet.

\subsubsection*{Type Ib/c Supernovae}

At central engine activity times $t_{\rm e}$ longer than the breakout time $t_{\rm b}$, the central engine activity time distribution is extremely steep, $p_{\rm e}\propto t_{\rm e}^{-\alpha}$ with $\alpha\sim 4.2$. Therefore, if this power-law distribution is extrapolated and assumed to be valid down to durations of $t_{\rm min}\sim 0.2\times t_{\rm b}$, the total rate of engines launching bipolar jets would be comparable to that of all Type Ib/c SNe. Hence it is tempting to conclude that $f_{\rm jet}\sim 1$ (i.e. most, or even all, SNe Ib/c have a central engine launching GRB-like jets), while only a small fraction ($f_{\rm out}\sim 10^{-2}$) of these jets manage to break out and power the prompt gamma-ray emission typical of GRBs. If this is the case, regular SNe would correspond to $t_{\rm e}\ll t_{\rm b}$, i.e. to jets choked long before the breakout.

Indeed, while all values in the range $0.01\lesssim f_{\rm jet}\lesssim 1$ are in principle possible, we argued throughout the paper that to avoid fine tuning of presumably unrelated quantities and in absence of further information the most natural value might be  $f_{\rm jet}~\simeq~1$. Such a scenario is also supported by (i) the increasing number of objects with intermediate properties between regular Type Ib/c SNe and long-soft GRBs (see for example \citealt{Margutti2014}); (ii) in particular, the existence of SNe without an associated GRB, which still show the signs of a jet's activity (e.g. \citealt{GranotRamirez2004, Mazzali2005, Paragi2010}). Our final physical picture is sketched in Figure \ref{fig:main}.

If most of Type Ib/c SNe have a GRB-like central engine, in the vast majority of cases the jets do not break out and instead deposit their energy into the star through $p\text{d}V$ work by the hot, high-pressure cocoon that they inflate while propagating inside the star at $t~<~t_{\rm e}~\ll~t_{\rm b}$. Interestingly, there are three fundamental energy scales which are comparable in all these events, namely: (i) the total energy in GRB jets, $E_{\rm jet}\sim 10^{49}\mhyphen10^{52}\text{ erg}$; (ii) the total energy released by CC SNe, $E_{\rm SN}\sim 10^{50}\mhyphen10^{52}\text{ erg}$; (iii) the binding energy (excluding the iron core) of pre-SN stars, $U_{\rm b}\sim 0.1\mhyphen2.5\times 10^{51}\text{ erg}$. Hence, jets that deposit at least part of their energy into the star may contribute significantly to the total energy associated with regular SN explosions. Moreover, if $f_{\rm jet}\sim 1$, this effect is potentially relevant for all Type Ib/c SNe. Such a contribution is particularly interesting since the more popular neutrino-driven mechanism faces several difficulties in reproducing the observations (see for example \citealt{Papish2015}).

A jet-driven scenario would naturally predict that all SNe explosions are intrinsically non-spherical due to the presence of a central engine launching bipolar jets, and that the observed asymmetry increases as $t_{\rm e}/t_{\rm b}$ approaches unity (or even exceeds it, in the rare cases when the jet breaks out and produces a GRB associated with the SN). This is in broad agreement with the trend for the asymmetry of the SNe explosions to increase with decreasing envelope mass and with the depth within the ejecta \citep{Wang2001}.

Our results have been confirmed by the subsequent analysis of \citet{Petropoulou2017}. These authors showed that the time distribution and the luminosity function of long-soft GRBs can be included into a coherent picture: the broken power-law luminosity function is due to the fact that less luminous jets are more easily chocked due to their longer breakout time (see the dependence of $t_{\rm b}$ on the jet's luminosity in Eq.~(\ref{eq:tb_ext})). They also found a long breakout time, consistent with an extended envelope surrounding the progenitors of long-soft GRBs, and showed the rate of central engines depositing $\sim 10^{51}\text{ erg}$ into the envelope to be comparable with that of Type Ib/c SNe.

\subsection{Future prospects}

Directions for future work include a detailed study of how different combinations of envelope properties (i.e. mass and radius) and engine durations affect (i) the light curve of the SN; (ii) their possible association to long-soft GRBs. Further investigation, possibly including the nucleosynthesis and/or the impact of the magnetic fields, is required to understand if the outlined scenario can actually induce at least part of the wide variety of observed events.

If this is the case, then studying the features of CC SNe explosions could be a unique opportunity to constrain the properties of GRB jets as well. For example, the study of the geometry of CC SNe explosions can also provide some hints about the jet's composition. In the context of Newtonian jets, \citet{Couch2009} realised that in jet-driven Type II SNe the thermal energy of the jet needs to dominate over the kinetic energy to avoid explosions that are much more asymmetric than inferred from observations. Hence, under the hypotheses that CC SNe have central engines launching bipolar jets, this result (derived for Newtonian jets) suggests that the geometry of the explosions may also provide important constraints on the composition of relativistic jets.

Different authors (e.g. \citealt{Fruchter2006,Svensson2010}) found that long GRBs are far more concentrated on the very brightest regions of their host galaxies than CC SNe, and the host galaxies of the long GRBs are significantly fainter and more irregular. Moreover, long GRBs prefer lower-metallicity hosts than broad-lined Type Ic SNe; this cannot be explained by the anti-correlation between star formation rate and metallicity, indicating a genuine aversion of the GRB progenitors towards metal-rich environments \citep{Modjaz2008b, GrahamFruchter2013}. Together these results suggest that long GRBs are associated with the most massive, metal-poor stars. If a significant fraction of Type Ib/c SNe have central engines launching relativistic jets, the fact that successful GRBs are biased towards the longest $t_{\rm e}$ (and shortest $t_{\rm b}$) could shed light on some fundamental physics, going beyond the purely phenomenological scheme adopted here. For example, one could still assume a narrow $t_{\rm b}$ distribution as we have done throughout this paper, and speculate that $t_{\rm e}$ is correlated (anti-correlated) with the mass (metallicity) of the progenitor star. Hence, the mass of the star would be one of the fundamental physical parameters shaping the activity time distribution of the central engines.

\section*{Acknowledgements}
ES is grateful to Yuri Lyubarsky for insightful discussions. The authors also thank Dovi Poznanski and John Magorrian for useful comments. ES and JG acknowledge support from the Israeli Science Foundation under Grant No. 719/14. OB is thankful for the support of the I-Core center of excellence of the CHE-ISF. MCS acknowledges support from the Deutsche Forschungsgemeinschaft in the Collaborative Research Center (SFB 881) ``The Milky Way System'' (subprojects B1, B2, and B8) and in the Priority Program SPP 1573 ``Physics of the Interstellar Medium'' (grant numbers KL 1358/18.1, KL 1358/19.2). MCS furthermore thanks the European Research Council for funding in the ERC Advanced Grant STARLIGHT (project number 339177).

%%%%%%%%%%%%%%%%%%%%%%%%%%%%%%%%%%%%%%%%%
\def\aap{A\&A}\def\aj{AJ}\def\apj{ApJ}\def\apjl{ApJ}\def\mnras{MNRAS}
\def\araa{ARA\&A}\def\physrep{PhR}\def\sovast{Sov. Astron.}\def\pasp{PASP}
\def\aapr{Astronomy \& Astrophysics Review}\def\apjs{ApJS}\def\nat{Nature}
\bibliographystyle{mn2e}
\bibliography{2d}

\begin{thebibliography}{79}
\expandafter\ifx\csname natexlab\endcsname\relax\def\natexlab#1{#1}\fi

\bibitem[{{Akiyama} {et~al}\mbox{.}(2003){Akiyama}, {Wheeler}, {Meier}, \&
  {Lichtenstadt}}]{Akiyama2003}
{Akiyama} S., {Wheeler} J.~C., {Meier} D.~L., {Lichtenstadt} I., 2003, \apj,
  584, 954

\bibitem[{{Bear} {et~al}\mbox{.}(2017){Bear}, {Grichener}, \&
  {Soker}}]{Bear2017}
{Bear} E., {Grichener} A., {Soker} N., 2017, arXiv:1706.00003

\bibitem[{{Bethe} \& {Wilson}(1985)}]{BetheWilson1985}
{Bethe} H.~A., {Wilson} J.~R., 1985, \apj, 295, 14

\bibitem[{{Bisnovatyi-Kogan}(1971)}]{Bisnovatyi-Kogan1971}
{Bisnovatyi-Kogan} G.~S., 1971, \sovast, 14, 652

\bibitem[{{Bloom} {et~al}\mbox{.}(2003){Bloom}, {Frail}, \&
  {Kulkarni}}]{Bloom2003}
{Bloom} J.~S., {Frail} D.~A., {Kulkarni} S.~R., 2003, \apj, 594, 674

\bibitem[{{Boissier} \& {Prantzos}(2009)}]{BoissierPrantzos2009}
{Boissier} S., {Prantzos} N., 2009, \aap, 503, 137

\bibitem[{{Bromberg} {et~al}\mbox{.}(2015){Bromberg}, {Granot}, \&
  {Piran}}]{Bromberg2015}
{Bromberg} O., {Granot} J., {Piran} T., 2015, \mnras, 450, 1077

\bibitem[{{Bromberg} {et~al}\mbox{.}(2011{\natexlab{a}}){Bromberg}, {Nakar}, \&
  {Piran}}]{Bromberg2011b}
{Bromberg} O., {Nakar} E., {Piran} T., 2011{\natexlab{a}}, \apjl, 739, L55

\bibitem[{{Bromberg} {et~al}\mbox{.}(2011{\natexlab{b}}){Bromberg}, {Nakar},
  {Piran}, \& {Sari}}]{Bromberg2011}
{Bromberg} O., {Nakar} E., {Piran} T., {Sari} R., 2011{\natexlab{b}}, \apj,
  740, 100

\bibitem[{{Bromberg} {et~al}\mbox{.}(2012){Bromberg}, {Nakar}, {Piran}, \&
  {Sari}}]{Bromberg2012}
{Bromberg} O., {Nakar} E., {Piran} T., {Sari} R., 2012, \apj, 749, 110

\bibitem[{{Bromberg} {et~al}\mbox{.}(2013){Bromberg}, {Nakar}, {Piran}, \&
  {Sari}}]{Bromberg2013}
{Bromberg} O., {Nakar} E., {Piran} T., {Sari} R., 2013, \apj, 764, 179

\bibitem[{{Burrows}(2013)}]{Burrows2013}
{Burrows} A., 2013, Reviews of Modern Physics, 85, 245

\bibitem[{{Butler} {et~al}\mbox{.}(2010){Butler}, {Bloom}, \&
  {Poznanski}}]{Butler2010}
{Butler} N.~R., {Bloom} J.~S., {Poznanski} D., 2010, \apj, 711, 495

\bibitem[{{Campana} {et~al}\mbox{.}(2006){Campana}, {Mangano}, {Blustin},
  {Brown}, {Burrows}, {Chincarini}, {Cummings}, {Cusumano}, {Della Valle},
  {Malesani}, {M{\'e}sz{\'a}ros}, {Nousek}, {Page}, {Sakamoto}, {Waxman},
  {Zhang}, {Dai}, {Gehrels}, {Immler}, {Marshall}, {Mason}, {Moretti},
  {O'Brien}, {Osborne}, {Page}, {Romano}, {Roming}, {Tagliaferri}, {Cominsky},
  {Giommi}, {Godet}, {Kennea}, {Krimm}, {Angelini}, {Barthelmy}, {Boyd},
  {Palmer}, {Wells}, \& {White}}]{Campana2006}
{Campana} S. {et~al.}, 2006, \nat, 442, 1008

\bibitem[{{Cano} {et~al}\mbox{.}(2017){Cano}, {Wang}, {Dai}, \&
  {Wu}}]{Cano2016}
{Cano} Z., {Wang} S.-Q., {Dai} Z.-G., {Wu} X.-F., 2017, Advances in Astronomy,
  2017, 8929054

\bibitem[{{Cappellaro} {et~al}\mbox{.}(2015){Cappellaro}, {Botticella},
  {Pignata}, {Grado}, {Greggio}, {Limatola}, {Vaccari}, {Baruffolo}, {Benetti},
  {Bufano}, {Capaccioli}, {Cascone}, {Covone}, {De Cicco}, {Falocco}, {Della
  Valle}, {Jarvis}, {Marchetti}, {Napolitano}, {Paolillo}, {Pastorello},
  {Radovich}, {Schipani}, {Spiro}, {Tomasella}, \& {Turatto}}]{Cappellaro2015}
{Cappellaro} E. {et~al.}, 2015, \aap, 584, A62

\bibitem[{{Chevalier}(2012)}]{Chevalier2012}
{Chevalier} R.~A., 2012, \apjl, 752, L2

\bibitem[{{Couch} {et~al}\mbox{.}(2009){Couch}, {Wheeler}, \&
  {Milosavljevi{\'c}}}]{Couch2009}
{Couch} S.~M., {Wheeler} J.~C., {Milosavljevi{\'c}} M., 2009, \apj, 696, 953

\bibitem[{{Crowther}(2007)}]{Crowther2007}
{Crowther} P.~A., 2007, \araa, 45, 177

\bibitem[{{Dahlen} {et~al}\mbox{.}(2012){Dahlen}, {Strolger}, {Riess},
  {Mattila}, {Kankare}, \& {Mobasher}}]{Dahlen2012}
{Dahlen} T., {Strolger} L.-G., {Riess} A.~G., {Mattila} S., {Kankare} E.,
  {Mobasher} B., 2012, \apj, 757, 70

\bibitem[{{Dahlen} {et~al}\mbox{.}(2004){Dahlen}, {Strolger}, {Riess},
  {Mobasher}, {Chary}, {Conselice}, {Ferguson}, {Fruchter}, {Giavalisco},
  {Livio}, {Madau}, {Panagia}, \& {Tonry}}]{Dahlen2004}
{Dahlen} T. {et~al.}, 2004, \apj, 613, 189

\bibitem[{{Drout} {et~al}\mbox{.}(2011){Drout}, {Soderberg}, {Gal-Yam},
  {Cenko}, {Fox}, {Leonard}, {Sand}, {Moon}, {Arcavi}, \& {Green}}]{Drout2011}
{Drout} M.~R. {et~al.}, 2011, \apj, 741, 97

\bibitem[{{Foreman-Mackey} {et~al}\mbox{.}(2013){Foreman-Mackey}, {Hogg},
  {Lang}, \& {Goodman}}]{emcee2013}
{Foreman-Mackey} D., {Hogg} D.~W., {Lang} D., {Goodman} J., 2013, \pasp, 125,
  306

\bibitem[{{Frail} {et~al}\mbox{.}(2001){Frail}, {Kulkarni}, {Sari},
  {Djorgovski}, {Bloom}, {Galama}, {Reichart}, {Berger}, {Harrison}, {Price},
  {Yost}, {Diercks}, {Goodrich}, \& {Chaffee}}]{Frail2001}
{Frail} D.~A. {et~al.}, 2001, \apjl, 562, L55

\bibitem[{{Friedman} \& {Bloom}(2005)}]{Friedman2005}
{Friedman} A.~S., {Bloom} J.~S., 2005, \apj, 627, 1

\bibitem[{{Fruchter} {et~al}\mbox{.}(2006){Fruchter}, {Levan}, {Strolger},
  {Vreeswijk}, {Thorsett}, {Bersier}, {Burud}, {Castro Cer{\'o}n},
  {Castro-Tirado}, {Conselice}, {Dahlen}, {Ferguson}, {Fynbo}, {Garnavich},
  {Gibbons}, {Gorosabel}, {Gull}, {Hjorth}, {Holland}, {Kouveliotou}, {Levay},
  {Livio}, {Metzger}, {Nugent}, {Petro}, {Pian}, {Rhoads}, {Riess}, {Sahu},
  {Smette}, {Tanvir}, {Wijers}, \& {Woosley}}]{Fruchter2006}
{Fruchter} A.~S. {et~al.}, 2006, \nat, 441, 463

\bibitem[{{Goldstein} {et~al}\mbox{.}(2016){Goldstein}, {Connaughton},
  {Briggs}, \& {Burns}}]{Goldstein2016}
{Goldstein} A., {Connaughton} V., {Briggs} M.~S., {Burns} E., 2016, \apj, 818,
  18

\bibitem[{{Goodman} \& {Weare}(2010)}]{GoodmanWeare2010}
{Goodman} J., {Weare} J., 2010, Comm. App. Math. Comp. Sci., 5, 65

\bibitem[{{Graham} \& {Fruchter}(2013)}]{GrahamFruchter2013}
{Graham} J.~F., {Fruchter} A.~S., 2013, \apj, 774, 119

\bibitem[{{Granot} \& {Ramirez-Ruiz}(2004)}]{GranotRamirez2004}
{Granot} J., {Ramirez-Ruiz} E., 2004, \apjl, 609, L9

\bibitem[{{Graur} {et~al}\mbox{.}(2011){Graur}, {Poznanski}, {Maoz}, {Yasuda},
  {Totani}, {Fukugita}, {Filippenko}, {Foley}, {Silverman}, {Gal-Yam},
  {Horesh}, \& {Jannuzi}}]{Graur2011}
{Graur} O. {et~al.}, 2011, \mnras, 417, 916

\bibitem[{{Guetta} \& {Della Valle}(2007)}]{GuettaDellavalle2007}
{Guetta} D., {Della Valle} M., 2007, \apjl, 657, L73

\bibitem[{{Guetta} {et~al}\mbox{.}(2005){Guetta}, {Piran}, \&
  {Waxman}}]{Guetta2005}
{Guetta} D., {Piran} T., {Waxman} E., 2005, \apj, 619, 412

\bibitem[{{Heger} {et~al}\mbox{.}(2004){Heger}, {Woosley}, {Langer}, \&
  {Spruit}}]{Heger2004}
{Heger} A., {Woosley} S.~E., {Langer} N., {Spruit} H.~C., 2004, in IAU
  Symposium, Vol. 215, Stellar Rotation, {Maeder} A., {Eenens} P., eds., p. 591

\bibitem[{{Hou} {et~al}\mbox{.}(2013){Hou}, {Liu}, {Lin}, {Wu}, \&
  {Lu}}]{Hou2013}
{Hou} S.-J., {Liu} T., {Lin} D.-B., {Wu} X.-F., {Lu} J.-F., 2013, in IAU
  Symposium, Vol. 290, Feeding Compact Objects: Accretion on All Scales,
  {Zhang} C.~M., {Belloni} T., {M{\'e}ndez} M., {Zhang} S.~N., eds., pp.
  223--224

\bibitem[{{Janka}(2012)}]{Janka2012}
{Janka} H.-T., 2012, Annual Review of Nuclear and Particle Science, 62, 407

\bibitem[{{Janka} {et~al}\mbox{.}(2016){Janka}, {Melson}, \&
  {Summa}}]{Janka2016}
{Janka} H.-T., {Melson} T., {Summa} A., 2016, Annual Review of Nuclear and
  Particle Science, 66, 341

\bibitem[{{Khokhlov} {et~al}\mbox{.}(1999){Khokhlov}, {H{\"o}flich}, {Oran},
  {Wheeler}, {Wang}, \& {Chtchelkanova}}]{Khokhlov1999}
{Khokhlov} A.~M., {H{\"o}flich} P.~A., {Oran} E.~S., {Wheeler} J.~C., {Wang}
  L., {Chtchelkanova} A.~Y., 1999, \apjl, 524, L107

\bibitem[{{Kouveliotou} {et~al}\mbox{.}(1993){Kouveliotou}, {Meegan},
  {Fishman}, {Bhat}, {Briggs}, {Koshut}, {Paciesas}, \&
  {Pendleton}}]{Kouveliotou1993}
{Kouveliotou} C., {Meegan} C.~A., {Fishman} G.~J., {Bhat} N.~P., {Briggs}
  M.~S., {Koshut} T.~M., {Paciesas} W.~S., {Pendleton} G.~N., 1993, \apjl, 413,
  L101

\bibitem[{{Kumar} \& {Zhang}(2015)}]{KumarZhang2015}
{Kumar} P., {Zhang} B., 2015, \physrep, 561, 1

\bibitem[{{Lazzati} {et~al}\mbox{.}(2012){Lazzati}, {Morsony}, {Blackwell}, \&
  {Begelman}}]{Lazzati2012}
{Lazzati} D., {Morsony} B.~J., {Blackwell} C.~H., {Begelman} M.~C., 2012, \apj,
  750, 68

\bibitem[{{Le} \& {Dermer}(2007)}]{LeDermer2007}
{Le} T., {Dermer} C.~D., 2007, \apj, 661, 394

\bibitem[{{LeBlanc} \& {Wilson}(1970)}]{LeblancWilson1970}
{LeBlanc} J.~M., {Wilson} J.~R., 1970, \apj, 161, 541

\bibitem[{{Li} {et~al}\mbox{.}(2011){Li}, {Chornock}, {Leaman}, {Filippenko},
  {Poznanski}, {Wang}, {Ganeshalingam}, \& {Mannucci}}]{Li2011}
{Li} W., {Chornock} R., {Leaman} J., {Filippenko} A.~V., {Poznanski} D., {Wang}
  X., {Ganeshalingam} M., {Mannucci} F., 2011, \mnras, 412, 1473

\bibitem[{{Liang} {et~al}\mbox{.}(2008){Liang}, {Racusin}, {Zhang}, {Zhang}, \&
  {Burrows}}]{Liang2008}
{Liang} E.-W., {Racusin} J.~L., {Zhang} B., {Zhang} B.-B., {Burrows} D.~N.,
  2008, \apj, 675, 528

\bibitem[{{MacFadyen} \& {Woosley}(1999)}]{MacfadyenWoosley1999}
{MacFadyen} A.~I., {Woosley} S.~E., 1999, \apj, 524, 262

\bibitem[{{MacFadyen} {et~al}\mbox{.}(2001){MacFadyen}, {Woosley}, \&
  {Heger}}]{Macfadyen2001}
{MacFadyen} A.~I., {Woosley} S.~E., {Heger} A., 2001, \apj, 550, 410

\bibitem[{{Maeda} {et~al}\mbox{.}(2008){Maeda}, {Kawabata}, {Mazzali},
  {Tanaka}, {Valenti}, {Nomoto}, {Hattori}, {Deng}, {Pian}, {Taubenberger},
  {Iye}, {Matheson}, {Filippenko}, {Aoki}, {Kosugi}, {Ohyama}, {Sasaki}, \&
  {Takata}}]{Maeda2008}
{Maeda} K. {et~al.}, 2008, Science, 319, 1220

\bibitem[{{Maeda} \& {Nomoto}(2003)}]{MaedaNomoto2003}
{Maeda} K., {Nomoto} K., 2003, \apj, 598, 1163

\bibitem[{{Margutti} {et~al}\mbox{.}(2015){Margutti}, {Guidorzi}, {Lazzati},
  {Milisavljevic}, {Kamble}, {Laskar}, {Parrent}, {Gehrels}, \&
  {Soderberg}}]{Margutti2015}
{Margutti} R. {et~al.}, 2015, \apj, 805, 159

\bibitem[{{Margutti} {et~al}\mbox{.}(2014){Margutti}, {Milisavljevic},
  {Soderberg}, {Guidorzi}, {Morsony}, {Sanders}, {Chakraborti}, {Ray},
  {Kamble}, {Drout}, {Parrent}, {Zauderer}, \& {Chomiuk}}]{Margutti2014}
{Margutti} R. {et~al.}, 2014, \apj, 797, 107

\bibitem[{{Mazzali} {et~al}\mbox{.}(2005){Mazzali}, {Kawabata}, {Maeda},
  {Nomoto}, {Filippenko}, {Ramirez-Ruiz}, {Benetti}, {Pian}, {Deng},
  {Tominaga}, {Ohyama}, {Iye}, {Foley}, {Matheson}, {Wang}, \&
  {Gal-Yam}}]{Mazzali2005}
{Mazzali} P.~A. {et~al.}, 2005, Science, 308, 1284

\bibitem[{{Melinder} {et~al}\mbox{.}(2012){Melinder}, {Dahlen}, {Menc{\'{\i}}a
  Trinchant}, {{\"O}stlin}, {Mattila}, {Sollerman}, {Fransson}, {Hayes},
  {Kankare}, \& {Nasoudi-Shoar}}]{Melinder2012}
{Melinder} J. {et~al.}, 2012, \aap, 545, A96

\bibitem[{{Modjaz} {et~al}\mbox{.}(2008{\natexlab{a}}){Modjaz}, {Kewley},
  {Kirshner}, {Stanek}, {Challis}, {Garnavich}, {Greene}, {Kelly}, \&
  {Prieto}}]{Modjaz2008b}
{Modjaz} M. {et~al.}, 2008{\natexlab{a}}, \aj, 135, 1136

\bibitem[{{Modjaz} {et~al}\mbox{.}(2008{\natexlab{b}}){Modjaz}, {Kirshner},
  {Blondin}, {Challis}, \& {Matheson}}]{Modjaz2008}
{Modjaz} M., {Kirshner} R.~P., {Blondin} S., {Challis} P., {Matheson} T.,
  2008{\natexlab{b}}, \apjl, 687, L9

\bibitem[{{Modjaz} {et~al}\mbox{.}(2016){Modjaz}, {Liu}, {Bianco}, \&
  {Graur}}]{Modjaz2016}
{Modjaz} M., {Liu} Y.~Q., {Bianco} F.~B., {Graur} O., 2016, \apj, 832, 108

\bibitem[{{Nakar}(2007)}]{Nakar2007}
{Nakar} E., 2007, \physrep, 442, 166

\bibitem[{{Nakar}(2015)}]{Nakar2015}
{Nakar} E., 2015, \apj, 807, 172

\bibitem[{{Nakar} \& {Sari}(2012)}]{NakarSari2012}
{Nakar} E., {Sari} R., 2012, \apj, 747, 88

\bibitem[{{Ostriker} \& {Gunn}(1971)}]{OstrikerGunn1971}
{Ostriker} J.~P., {Gunn} J.~E., 1971, \apjl, 164, L95

\bibitem[{{Papish} {et~al}\mbox{.}(2015){Papish}, {Nordhaus}, \&
  {Soker}}]{Papish2015}
{Papish} O., {Nordhaus} J., {Soker} N., 2015, \mnras, 448, 2362

\bibitem[{{Paragi} {et~al}\mbox{.}(2010){Paragi}, {Taylor}, {Kouveliotou},
  {Granot}, {Ramirez-Ruiz}, {Bietenholz}, {van der Horst}, {Pidopryhora}, {van
  Langevelde}, {Garrett}, {Szomoru}, {Argo}, {Bourke}, \&
  {Paczy{\'n}ski}}]{Paragi2010}
{Paragi} Z. {et~al.}, 2010, \nat, 463, 516

\bibitem[{{Petropoulou} {et~al}\mbox{.}(2017){Petropoulou}, {Barniol Duran}, \&
  {Giannios}}]{Petropoulou2017}
{Petropoulou} M., {Barniol Duran} R., {Giannios} D., 2017, ArXiv:1707.01914

\bibitem[{{Petrushevska} {et~al}\mbox{.}(2016){Petrushevska}, {Amanullah},
  {Goobar}, {Fabbro}, {Johansson}, {Kjellsson}, {Lidman}, {Paech}, {Richard},
  {Dahle}, {Ferretti}, {Kneib}, {Limousin}, {Nordin}, \&
  {Stanishev}}]{Petrushevska2016}
{Petrushevska} T. {et~al.}, 2016, \aap, 594, A54

\bibitem[{{Piran} {et~al}\mbox{.}(2017){Piran}, {Nakar}, {Mazzali}, \&
  {Pian}}]{Piran2017}
{Piran} T., {Nakar} E., {Mazzali} P., {Pian} E., 2017, ArXiv:1704.08298

\bibitem[{{Racusin} {et~al}\mbox{.}(2009){Racusin}, {Liang}, {Burrows},
  {Falcone}, {Sakamoto}, {Zhang}, {Zhang}, {Evans}, \& {Osborne}}]{Racusin2009}
{Racusin} J.~L. {et~al.}, 2009, \apj, 698, 43

\bibitem[{{Robertson} \& {Ellis}(2012)}]{RobertsonEllis2012}
{Robertson} B.~E., {Ellis} R.~S., 2012, \apj, 744, 95

\bibitem[{{Smartt}(2009)}]{Smartt2009}
{Smartt} S.~J., 2009, \araa, 47, 63

\bibitem[{{Smith} {et~al}\mbox{.}(2012){Smith}, {Cenko}, {Butler}, {Bloom},
  {Kasliwal}, {Horesh}, {Kulkarni}, {Law}, {Nugent}, {Ofek}, {Poznanski},
  {Quimby}, {Sesar}, {Ben-Ami}, {Arcavi}, {Gal-Yam}, {Polishook}, {Xu},
  {Yaron}, {Frail}, \& {Sullivan}}]{Smith2012}
{Smith} N. {et~al.}, 2012, \mnras, 420, 1135

\bibitem[{{Soderberg} {et~al}\mbox{.}(2006){Soderberg}, {Kulkarni}, {Nakar},
  {Berger}, {Cameron}, {Fox}, {Frail}, {Gal-Yam}, {Sari}, {Cenko}, {Kasliwal},
  {Chevalier}, {Piran}, {Price}, {Schmidt}, {Pooley}, {Moon}, {Penprase},
  {Ofek}, {Rau}, {Gehrels}, {Nousek}, {Burrows}, {Persson}, \&
  {McCarthy}}]{Soderberg2006}
{Soderberg} A.~M. {et~al.}, 2006, \nat, 442, 1014

\bibitem[{{Strolger} {et~al}\mbox{.}(2015){Strolger}, {Dahlen}, {Rodney},
  {Graur}, {Riess}, {McCully}, {Ravindranath}, {Mobasher}, \&
  {Shahady}}]{Strolger2015}
{Strolger} L.-G. {et~al.}, 2015, \apj, 813, 93

\bibitem[{{Svensson} {et~al}\mbox{.}(2010){Svensson}, {Levan}, {Tanvir},
  {Fruchter}, \& {Strolger}}]{Svensson2010}
{Svensson} K.~M., {Levan} A.~J., {Tanvir} N.~R., {Fruchter} A.~S., {Strolger}
  L.-G., 2010, \mnras, 405, 57

\bibitem[{{Taubenberger} {et~al}\mbox{.}(2009){Taubenberger}, {Valenti},
  {Benetti}, {Cappellaro}, {Della Valle}, {Elias-Rosa}, {Hachinger},
  {Hillebrandt}, {Maeda}, {Mazzali}, {Pastorello}, {Patat}, {Sim}, \&
  {Turatto}}]{Taubenberger2009}
{Taubenberger} S. {et~al.}, 2009, \mnras, 397, 677

\bibitem[{{Wanderman} \& {Piran}(2010)}]{WandermanPiran2010}
{Wanderman} D., {Piran} T., 2010, \mnras, 406, 1944

\bibitem[{{Wang} {et~al}\mbox{.}(2001){Wang}, {Howell}, {H{\"o}flich}, \&
  {Wheeler}}]{Wang2001}
{Wang} L., {Howell} D.~A., {H{\"o}flich} P., {Wheeler} J.~C., 2001, \apj, 550,
  1030

\bibitem[{{Wang} \& {Wheeler}(2008)}]{WangWheeler2008}
{Wang} L., {Wheeler} J.~C., 2008, \araa, 46, 433

\bibitem[{{Woosley} \& {Bloom}(2006)}]{WoosleyBloom2006}
{Woosley} S.~E., {Bloom} J.~S., 2006, \araa, 44, 507

\bibitem[{{Woosley} {et~al}\mbox{.}(2002){Woosley}, {Heger}, \&
  {Weaver}}]{Woosley2002}
{Woosley} S.~E., {Heger} A., {Weaver} T.~A., 2002, Reviews of Modern Physics,
  74, 1015

\bibitem[{{Yaron} {et~al}\mbox{.}(2017){Yaron}, {Perley}, {Gal-Yam}, {Groh},
  {Horesh}, {Ofek}, {Kulkarni}, {Sollerman}, {Fransson}, {Rubin}, {Szabo},
  {Sapir}, {Taddia}, {Cenko}, {Valenti}, {Arcavi}, {Howell}, {Kasliwal},
  {Vreeswijk}, {Khazov}, {Fox}, {Cao}, {Gnat}, {Kelly}, {Nugent}, {Filippenko},
  {Laher}, {Wozniak}, {Lee}, {Rebbapragada}, {Maguire}, {Sullivan}, \&
  {Soumagnac}}]{Yaron2017}
{Yaron} O. {et~al.}, 2017, ArXiv:1701.02596

\end{thebibliography}

\appendix
\section{Fitting the GRB time distribution}
\label{sec:appendix}

To attack the problem we model the time distribution of collapsars through our Eq. \eqref{eq:Pe1}, while we assume a log-normal distribution for non-collapsar objects. Specifically, we use
\begin{equation}
\label{eq:t_distr}
P_{\rm \gamma,tot}=f_{\rm nc}P_{\rm \gamma,nc}+\left(1-f_{\rm nc}\right)P_{\rm \gamma,c} \;,
\end{equation}
where $f_{\rm nc}$ is the fraction of non-collapsar objects in the sample and
\begin{align}
\label{eq:Pnc}
P_{\rm \gamma,nc}\left(T_\gamma\right)&=\frac{1}{T_\gamma\sigma\sqrt{2\pi}}\e^{-\frac{\left(\ln T_\gamma-\mu\right)^2}{2\sigma^2}}\;,\\
\label{eq:Pc}
P_{\rm \gamma,c}\left(T_\gamma\right) & =\frac{\alpha-1}{\hat{T}_{\rm b}}
\left(1+\frac{T_\gamma}{\hat{T}_{\rm b}}\right)^{-\alpha}\;,
\end{align}
are the separate contributions of non-collapsars and collapsars respectively. Hence, we eventually fit the observed GRB time distribution with five parameters (note that we are using one parameter less than previous models; e.g. \citealt{Bromberg2013}).

Our job is to constrain the five free parameters of our model ($M$) based on the data ($D$). In particular, we seek the posterior probability distribution $P\left(M|D\right)$. By Bayes' theorem, $P\left(M|D\right)$ is proportional to $P\left(D|M\right)\times P\left(M\right)$, where $P\left(M\right)$ is our prior on $M$. Here we use flat priors on all the parameters.\footnote{We checked that the choice of the priors has a small impact on the results. For example, using flat priors on the logarithms of the parameters changes their best fit values by $\lesssim 5\%$, which is well within uncertainties.}
The likelihood $P\left(D|M\right)$ can be calculated as
\begin{equation}
\log\left[P\left(D|M\right)\right]=\sum_{\text{GRB}}\log\left[P_{\rm \gamma,tot}\left(T_\gamma\right)\right]\;,
\end{equation}
where we sum over all the measured GRB durations. The  observed  GRB  duration $T_\gamma$ is taken to be $T_{90}$,  i.e. the time over which the central $90\%$ of the photon counts from the GRB are measured.

The main advantages of using a Bayesian approach are that (i) it results in a complete understanding of the posterior probability distribution; (ii) the fit is independent of any (arbitrary) binning of the data.

Specifically, we fit our model to the data using a Markov-Chain Monte-Carlo (MCMC) simulation. We use the affine invariant ensemble sampler developed by
\citet{GoodmanWeare2010}. This algorithm has been modified and improved before being released as the publicly available Python module {\small EMCEE} by \citet{emcee2013}.

The value of the parameters and their uncertainties (associated with $68\%$ probability contours of the marginalised probability distributions) are given in Table \ref{table:fit}. In particular, we predict a fraction $f_{\rm nc}= 7\pm1\%$ of the GRBs in the {\it Swift} sample to be non-collapsars. According to our model, there is a $50\%$ probability for a GRB to be have a collapsar origin when $T_\gamma\approx 0.8\;$s, in agreement with previous results for {\it Swift} GRBs \citep{Bromberg2013}.

We checked {\it a posteriori} the validity of our fit using the $\chi^2$ test. Only at this point, the binning of the data is required. For simplicity we use equally spaced logarithmic bins, with $\Delta\log_{10}\left(T_\gamma\right)~=~0.2$. Assuming a poissonian variance in each bin, we find a $\chi^2/\text{DOF}~=~1.3$ with $16$ Degrees Of Freedom, corresponding to a $p$-value $p~=~0.16$. The size of the bins does not change the global result of the test (we typically find $\chi^2/\text{DOF}~\approx~1.0\mhyphen1.5$).

\begin{table}
\caption{Best fit parameters for the GRB time distribution.}
\begin{tabular} {c c c c c}
\toprule
$\alpha$ & $\hat{T}_{\rm b}\;$(s) & $f_{\rm nc}$ & $\mu$ & $\sigma$ \\
\midrule
$4.2_{-0.5}^{+0.6}$ & $170_{-30}^{+40}$ & $0.07_{-0.01}^{+0.01}$ & $-1.3_{-0.2}^{+0.3}$ & $1.0_{-0.2}^{+0.3}$\\
\bottomrule
\end{tabular}
\label{table:fit}
\end{table}

\subsubsection*{Selection effects}

In principle, since longer GRBs are generally less luminous, one may wonder whether the decline of the GRB duration distribution at $T_\gamma\gtrsim \hat{T}_{\rm b}$ is partially due to the sensitivity of the detector.
In particular, some of the longest GRBs may be either (i) missing, because they are too faint to be detected, or (ii) classified as shorter ones, since only part of their light curve is luminous enough (e.g. some early/late pulses may be missing or too dim to be included in $T_{90}$). Below we discuss these two effects in turn.

The first effect is mitigated by the fact that the longest GRBs emit most of their radiation during a small fraction (typically $\lesssim 5\%$; \citealt{Butler2010}) of the activity time. Hence, most of them manage to trigger the {\it Swift} BAT detector and the final correction is likely modest (\citealt{Butler2010}, see their Figure 2).

The second effect is harder to quantify. Since it moves GRBs from longer to shorter durations, then as a tentative approach we model the effect of the detector sensitivity by an exponential cutoff, $\exp\left(-T_\gamma/T_{\rm cutoff}\right)$, on the time distribution of collapsar GRBs, $P_{\rm \gamma,c}\left(T_\gamma\right)$. Note that, since Eq. \eqref{eq:Pc} should be still normalised to unity, the predicted number of GRBs with $T_\gamma\ll T_{\rm cutoff}$ increases (so this indeed mimics the effect of moving GRBs from longer to shorter durations). In general, the fit becomes worse while decreasing $T_{\rm cutoff}$, and already when $T_{\rm cutoff}\sim 800\text{ s}$ we find $\chi^2/\text{DOF}\sim 3$ for the best fit. Since the relevant parameters change by $\lesssim 10\%$ (well within error bars) for $T_{\rm cutoff}\gtrsim 800\text{ s}$, we expect our final results to be fairly robust against these uncertainties.

\section{On the possible extension to Type II SN\MakeLowercase{e}}
\label{sec:typeII}

Our constraints on $f_{\rm jet}$ are strictly valid only for the Type Ib/c SNe considered so far. However, in the context of the collapsar model for long GRBs, it was soon realised that the lack of hydrogen is necessary to avoid most of the jet's energy being dissipated well within the star \citep{MacfadyenWoosley1999}. Hence, it may seem natural to extend our results to the entire class of CC SNe, including Type II. If this is actually the case, then the presence of a larger envelope (corresponding to a breakout time significantly longer than what we estimate here) would be the only reason why we are not observing any Type II SN associated with a GRB. Interestingly, the possibility for some Type II SNe to be engine-driven has been explored both observationally (e.g. for the SN 2010jp; \citealt{Smith2012}) and theoretically (e.g. \citealt{Chevalier2012}). Finally, the fact that Type II SNe may be surrounded by dense circumstellar material (e.g. \citealt{Yaron2017}) further strengthen this conjecture.

Note that our main argument, which is based on the ratio of the GRB to the Type Ib/c SNe rates, can be extended with small modifications to include the entire CC SNe sample. For example, using the rate of all CC SNe (instead of Type Ib/c only) the result of Eq. \eqref{eq:tmin2} with $f_{\rm jet}=1$ becomes $t_{\rm min}/\hat{t}_{\rm b} = 0.17_{-0.06}^{+0.11}$.

However, the assumption that $f_{\rm jet}\sim 1$ also for Type II SNe raises a few questions. Based on our current state of knowledge launching a powerful enough jet seems to require (i) a rapid rotation and (ii) strong magnetic fields. In principle, the magnetorotational instability could provide large magnetic fields quite ubiquitously in CC SNe explosions (see for example \citealt{Akiyama2003}). However, for this mechanism to result in jets strong enough to help powering the SN explosion, a very fast rotation of the stellar iron core is required, in which case the alpha-omega dynamo may dominate the magnetic field amplification. This is in tension with the general predictions of stellar evolution models (e.g. \citealt{Heger2004}).\footnote{Note that these models usually focus on Type II SNe, while the evolution of Type Ib/c progenitors, which likely involve a strongly-interacting binary stellar system, is more uncertain (see for example \citealt{Smartt2009}).}

Hence, at the moment it seems unlikely for jets to be present in most Type II SNe. However, nature shows that the presence of jets is, quite unexplainably, ubiquitous in astrophysical objects, and their launching mechanism is still largely unclear. We might be surprised once again. In the following we will sometimes refer to the broad family of CC SNe. However, one should keep in mind that our results strictly apply only to Type Ib/c SNe.

\section{Propagation of the jet in a low-mass, extended envelope}
\label{sec:appendix2}

If the GRB's progenitor star is surrounded by a low-mass, extended envelope, then the propagation through such an envelope would dominate the breakout time of the relativistic jet launched by the central engine. While in the inner parts of the star the jet is collimated (i.e. cylindrical) and its head is at most mildly relativistic, when the jet enters the low-density, extended envelope it becomes uncollimated (i.e. conical) and propagates relativistically.

To calculate the breakout time we follow the framework developed by \citet{Bromberg2011}. The velocity of the jet's head through the star is
\begin{equation}
\label{eq:beta}
\beta_{\rm h}=\left(1+\tilde{L}^{-1/2}\right)^{-1}\;,
\end{equation}
where $\tilde{L}=L_{\rm jet}/\Sigma_{\rm jet}\rho_{\rm ext}c^3$. Here $\rho_{\rm ext}$ is the density of the envelope just in front of the jet's head, while $L_{\rm jet}$  and $\Sigma_{\rm jet}$ are the luminosity and the cross section, respectively, of a symmetric, double-sided jet (note that $\tilde{L}\propto L_{\rm jet}/\Sigma_{\rm jet}$ is the same for a one-sided jet). This corresponds to a breakout time of
\begin{equation}
t_{\rm b}=\int_0^{\rm R_{\rm ext}}\frac{\text{d}z}{\beta_{\rm h}c}\left(1-\beta_{\rm h}\right)=\int_0^{\rm R_{\rm ext}}\frac{\text{d}z}{c}\tilde{L}^{-1/2}\;.
\end{equation}

The jet propagating in the envelope can be approximated as conical, and to maintain a constant half-opening angle $\theta_{\rm jet}$. In this case the cross section holds $\Sigma_{\rm jet}=2\pi\theta_{\rm jet}^2\,z^2$, where $z$ is the distance from the star's centre. Assuming a power-law density profile ($\rho_{\rm ext}~\propto~z^{-\alpha}$) we get a breakout time for the envelope
\begin{equation}
t_{\rm b}=
67f(\alpha)\left(\frac{L_{\rm jet,iso}}{10^{51}\text{ erg s$^{-1}$}}\right)^{-1/2} \left(\frac{R_{\rm ext}}{10^{13}\text{\,cm}}\right)^{1/2} \left(\frac{M_{\rm ext}}{0.01M_\odot}\right)^{1/2}\text{\;s}\;,
\end{equation}
where $f(\alpha)=\frac{4}{4-\alpha}(\frac{3-\alpha}{3})^{1/2}$, and $R_{\rm ext}$ ($M_{\rm ext}$) is the envelope's radius (mass). The numeric coefficient is given for $f(0)=1$, but it changes by $\lesssim 15$\% in the range $0<\alpha<2$ ($77\;$s for $\alpha=2$).

\end{document}